\documentclass[%
 reprint,
 amsmath,amssymb,
 aps,
]{revtex4-1}

\newcommand{\beginsupplement}{%
        \setcounter{table}{0}
        \renewcommand{\thetable}{S\arabic{table}}%
        \setcounter{figure}{0}
        \renewcommand{\thefigure}{S\arabic{figure}}%
     }

\usepackage{tabu}
\usepackage{color}
\usepackage{graphicx}
\usepackage{dcolumn}
\usepackage{bm}
\usepackage{hyperref}

\begin{document}

\preprint{APS/123-QED}

\title{A positive mood, a flexible brain}

\author{Richard F. Betzel$^1$}
\author{Theodore D. Satterthwaite$^2$}
\author{Joshua I. Gold$^3$}
\author{Danielle S. Bassett$^{1,4}$}
 \email{dsb @ seas.upenn.edu}
\affiliation{
 Department of Bioengineering, University of Pennsylvania, Philadelphia, PA, 19104
}
\affiliation{
 Neuropsychiatry Section, Department of Psychiatry, University of Pennsylvania, Philadelphia, PA, 19104
}
\affiliation{
 Department of Neuroscience, University of Pennsylvania, Philadelphia, PA, 19104
}
\affiliation{
Department of Electrical and Systems Engineering, University of Pennsylvania, Philadelphia, PA, 19104
}

\date{\today}

\begin{abstract}
Flexible reconfiguration of human brain networks supports cognitive flexibility and learning. However, modulating flexibility to enhance learning requires an understanding of the relationship between flexibility and brain state. In an unprecedented longitudinal data set, we investigate the relationship between flexibility and mood, demonstrating that flexibility is positively correlated with emotional state. Our results inform the modulation of brain state to enhance response to training in health and injury.
\end{abstract}

\maketitle


\begin{figure*}[t]
\begin{center}
\centerline{\includegraphics[width=1\textwidth]{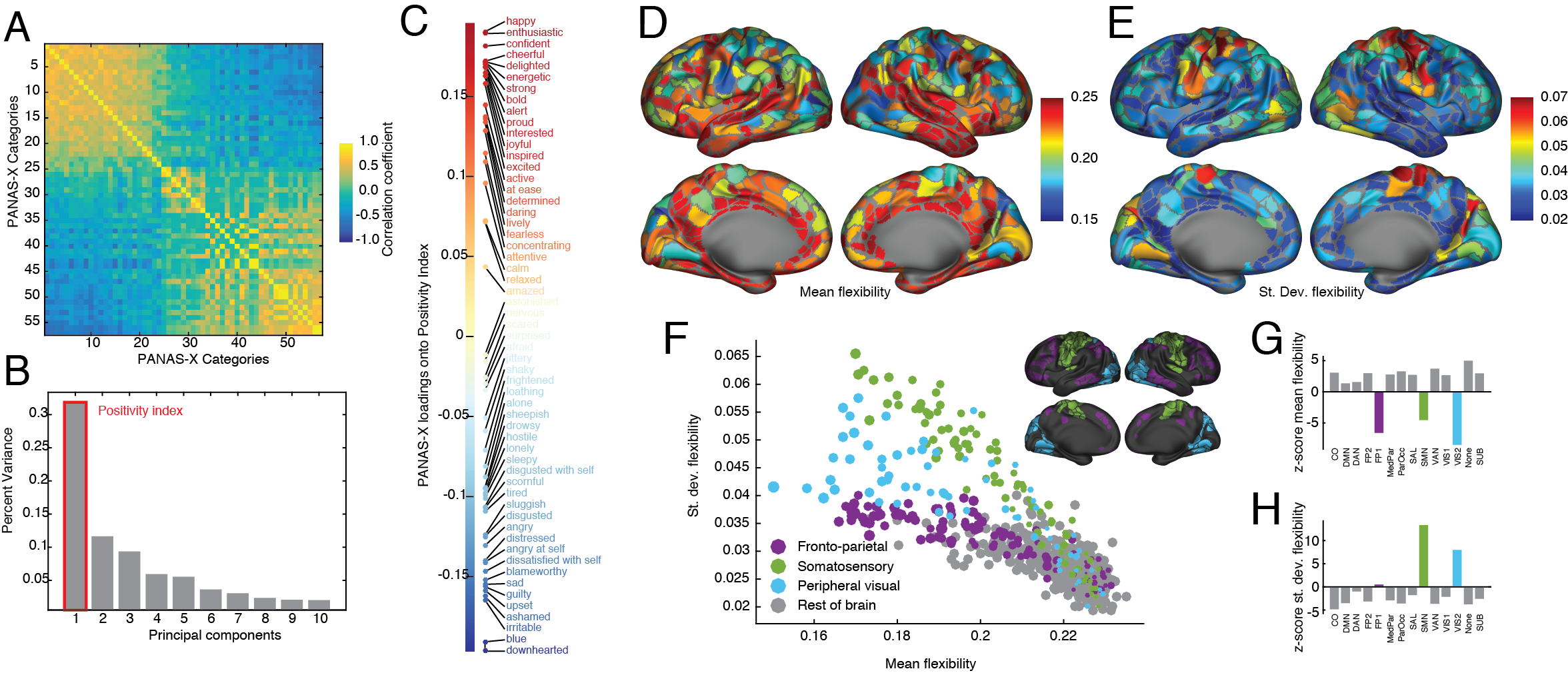}}
\caption{\textbf{Summary of principal component and flexibility analyses.} \emph{(A)} Correlation matrix of PANAS-X categories with one another, ordered by loading of PANAS-X categories onto the first principal component. \emph{(B)} Variance accounted for by each of the first ten principal components. \emph{(C)} Detail of the positivity index. Each colored point indicates the loading of a PANAS-X category onto the first principal component, which we termed the ``positivity index.'' \emph{(D,E)} Topographic representation of mean and standard deviation of flexibility. \emph{(F)} Mean and standard deviation of flexibility scores plotted against one another. Color-coding indicates brain systems: fronto-parietal (purple), somatomotor (green), peripheral visual (blue), and the rest of brain (grey). Inset depicts topographic representation of somatomotor (SMN), fronto-parietal (FP1), and peripheral visual (VIS2) \emph{(G)} The $z$-score of the mean system flexibility. \emph{(H)} The $z$-score of the standard deviation of system flexibility.}\label{figure:fig1}
\end{center}
\end{figure*}

Transient changes in the patterns of communication between brain areas support both learning and cognitive flexibility \cite{bassett2011dynamic,braun2015dynamic}. These abilities are not static but can vary considerably over time and as a function of an individual's affective state. For example, learning often shows an ``inverted-U'' relationship with arousal, with optimal learning at moderate levels of arousal \cite{robert1908relation}. Together, these findings imply that the influence of affective state on learning and cognition may involve modulations of brain network flexibility. However, virtually nothing is known about such modulations.

A potential simple and intuitive affect-related driver of daily variations in brain network flexibility is mood \cite{critchley2005neural}. Mood can fluctuate normally over time scales ranging from minutes to weeks. Moreover, mood can affect learning, for example by biasing the perception of reward outcomes \cite{eldar2016mood}. These biases are thought to arise from neurophysiological changes in neurotransmitter systems linked to arousal \cite{nassar2012rational}. However, the network-level mechanisms of these processes in the human brain remain unknown. We hypothesized that positive mood is associated with enhanced brain network flexibility, potentially explaining the observations that more flexible brains display greater cognitive flexibility and better learning \cite{bassett2011dynamic,braun2015dynamic}.

To address this hypothesis, we leveraged data from the \textit{MyConnectome Project}, which acquired extensive longitudinal neuroimaging and psychophysiological data from a single participant \cite{laumann2015functional, poldrack2015long}, who underwent multiple resting-state fMRI scans each week for a year. The participant also recorded his mood on a standard questionnaire (expanded Positive and Negative Affect Schedule; PANAS-X) \cite{watson1988development}. Subjective ratings across the 60 mood categories were correlated with one another, suggesting the presence of a latent structure (Table \ref{table:panasxTerms}; Figure~\ref{figure:fig1}A). We interrogated this structure using a principal components analysis, generating a set of mutually orthogonal components, loadings of PANAS-X categories onto components, and the percent variance accounted for by each component. The first component explained $\approx$33\% of the variance (the next accounted for $\approx$12\%) (Figure \ref{figure:fig1}B). The top three PANAS-X categories, in terms of their loading magnitude onto the first component, were ``happy'', ``enthusiastic'', and ``confident''. The bottom three were ``downhearted'', ``blue'', and ``irritable''. We termed the first principal component ``positivity index'' ($PI$), due to its sensitivity to emotional valence (Figure \ref{figure:fig1}C).

Next, we estimated flexibility in 630 brain regions and on average across the brain. This entailed dividing regional fMRI BOLD time series into non-overlapping windows. Within each window, we estimated functional connectivity between all pairs of brain regions using a magnitude-squared wavelet coherence \cite{bassett2011dynamic}. The result was an ordered set of functional connectivity matrices, each of which represented a layer in a multi-layer network \cite{bassett2013robust}. Next, we used a community detection algorithm to partition brain regions into communities (functional sub-systems \cite{power2011functional}) across layers (windows) \cite{mucha2010community} (Figures~\ref{figure:exampleLouvainOutput},\ref{figure:paramSweep}). Using these community assignments, we calculated the flexibility of each brain region as the fraction of times that its community assignment changed from one layer to the next \cite{bassett2011dynamic}. We repeated this analysis for all scan sessions.

We first asked which regions of the brain were flexible \emph{versus} inflexible, and which region's flexibility values varied appreciably across scan sessions. We observed that most brain regions possessed similar levels of mean flexibility and quotidian variability; the latter measured by the standard deviation (Figures~\ref{figure:fig1}D,E). A small number of regions, however, including components of visual, fronto-parietal, and somatomotor systems, possessed lower mean flexibility than the rest of the brain and were also more variable (Figure~\ref{figure:fig1}F). To quantify these observations, we aggregated regional flexibility by brain system and found that these same systems had mean flexibilities much lower than expected by chance (permutation test, $z_{FP2} = -6.52$, $p = 3.44 \times 10^{-11}$; $z_{SMN} = -4.53$, $p = 3.00 \times 10^{-6}$; $z_{VIS2} = -8.52$, $p<10^{-15}$; FDR-controlled, $d = 0.001$) (Figure~\ref{figure:fig1}G). Similarly, the quotidian variability of SMN and VIS2 were much greater than expected (permutation test, $z_{SMN} = 13.45$, $p<10^{-15}$; $z_{VIS2} = 7.93$, $p<10^{-15}$; FDR-controlled, $d = 0.001$) (Figure~\ref{figure:fig1}H). The presence of high quotidian variability in relatively rigid regions suggests the presence of a strong energetic constraint on network dynamics.

Global flexibility, or the average flexibility across brain regions, was positively correlated with positivity index (Pearson's correlation, $\hat{r}(PI,F) = 0.289, p = 0.013$) (Figure \ref{figure:fig2}A) (See Supplementary Information for details; Figures~\ref{figure:leaveOneOut}--\ref{figure:nuisance}), implying that positive emotional states correspond to a more flexible brain. To better understand which brain regions contributed to this correlation, we calculated the correlation of each brain region's flexibility with the positivity index. We observed that most brain regions were not significantly associated with the positivity index. However, regions comprising the somatomotor system exhibited significant positive correlations (permutation test, $z = 8.40, p < 10^{-15}$) (Figure \ref{figure:fig2}B,C), complementing prior work linking heightened motor activations -- potentially due to motor imagery -- with positive mood \cite{adolphs2002neural}.  A few systems were also more anti-correlated with flexibility than expected, including cingulo-opercular ($z_{CO} = -3.38$, $p = 3.63 \times 10^{-4}$), dorsal attention ($z_{DAN} = -4.14$, $p = 1.66 \times 10^{-5}$), and peripheral visual networks ($z_{VIS2} = -3.52$, $p = 2.11 \times 10^{-4}$) (FDR-controlled, $d = 0.001$), suggesting that relative stability in these higher-order cognitive systems was accompanied by positive mood.
\begin{figure*}[t]
\begin{center}
\centerline{\includegraphics[width=1\textwidth]{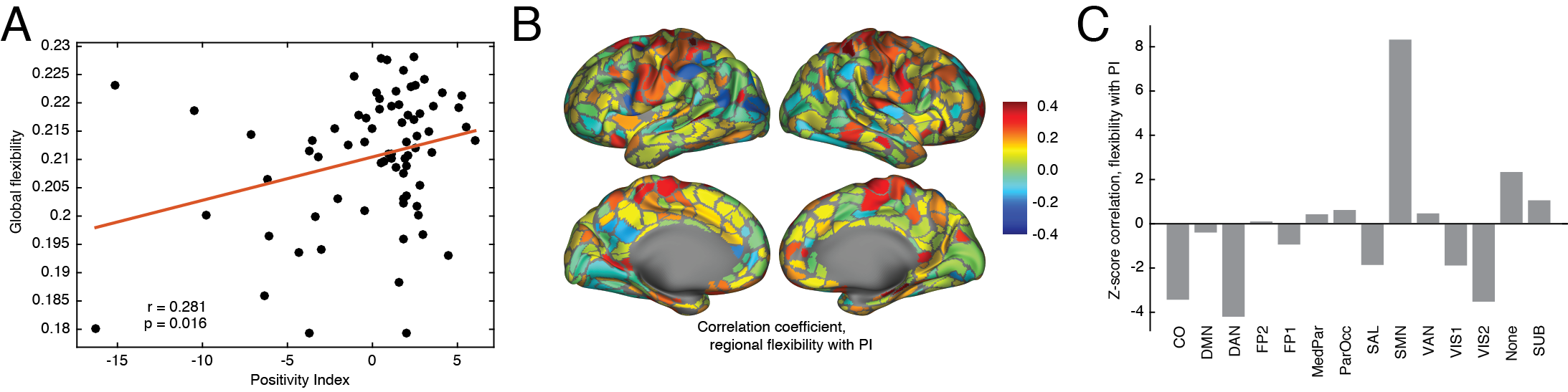}}
\caption{\textbf{(Relationship of flexibility scores with positivity index.)} \emph{(A)} Global flexibility as a function of positivity index. \emph{(B)} Topographic representation of correlation of regional flexibility with positivity index. \emph{(C)} Regional correlations grouped by system between regional flexibility and positivity index.}\label{figure:fig2}
\end{center}
\end{figure*}

The relationship between mood and flexibility suggests a potential network-level mechanism for learning deficits observed in mood disorders \cite{chrobak2015implicit}, and the dependence of those deficits on cognitive flexibility \cite{papmeyer2015neurocognition}. In these individuals, the development of pharmacological and stimulation-based interventions to alter brain network flexibility is therefore of particular interest. For example, brain network flexibility can be altered through modulation of NMDA receptor function \cite{braun2016dynamic}. However, an arguably more powerful approach might be to target states of arousal, which are known to be altered in mood disorders \cite{hegerl2014vigilance}, implicated in learning \cite{critchley2005neural}, and modulated by norepinephrine systems \cite{critchley2005neural}.  There is some preliminary evidence that arousal modulates network connectivity \cite{eldar2016mood}, but further work is needed to understand the patterns and dynamics of these network modulations and their relationship to mood.

Neurophysiological underpinnings aside, our observations can nevertheless inform the development of educational interventions to enhance learning. Intuitively, our results support the notion that by altering mood, one might alter brain network flexibility, and therefore predispose the brain to learn quickly in subsequent tasks. Such an outcome would directly fulfill the goals of personalized neuroeducation \cite{devonshire2010neuroscience}: the use of neuroscientific information to inform educational practices tuned to individual students.  Potentially powerful modulations could include simple mental exercises, which are easily translated into educational settings. For example, self-affirmation tasks have been shown to parametrically alter brain activity \cite{cascio2015self} to a degree that predicts individual differences in future behavior \cite{falk2015self}. Future work could define a carefully titrated library of mental tasks that modulate brain network flexibility (and subsequent learning) in a predictable fashion by modulating mood.

\section*{Online methods}
\subsection*{MyConnectome data}
All data were obtained from the \textit{MyConnectome Project}'s data-sharing webpage (\url{http://myconnectome.org/wp/data-sharing/}). Specifically, we studied pre-processed parcel fMRI time series for scan sessions 14--104. Details of the pre-processing procedure have been described elsewhere \cite{laumann2015functional, poldrack2015long}. Each session consisted of 518 time points during which the average fMRI BOLD signal was measured for $N=630$ parcels or regions of interest (ROIs). With a TR of 2.2 s, the analyzed segment of each session was approximately 19 minutes long. In addition to fMRI data, we also examined behavioral data available on the same webpage. The behavioral data included additional biometric information, such as heart rate and blood pressure, though we analyzed only PANAS-X categories, a set of emotional categories that the subject rated on a 0-5 Likert scale. Usually the PANAS-X test includes 60 categories. Only 57 were used as part of our analysis; the categories ``bashful'', ``timid'', and ``shy'' had ratings of zero for the entire duration of the experiment.

\subsection*{Principal component analysis}
Our analysis focused on the $n=73$ scan sessions for which both resting-state fMRI data and all $p=57$ PANAS-X categories were available. We standardized each category to have zero mean and unit variance. We represented the full set as the matrix, $\mathbf{X} \in \mathbb{R}^{n \times p}$, which we submitted to a principal component analysis (PCA). Essentially, PCA takes a data matrix and linearly factorizes it by creating a set of orthogonal \textit{principal components}, subject to the condition that each successive component has the greatest possible variance. Each component is a linear combination of the original data variables.

Specifically, we performed PCA using a singular value decomposition (SVD) \cite{eckart1936approximation} which deconstructs $\mathbf{X}$ according to the equation:

\begin{equation}
\mathbf{X} = \mathbf{U} \mathbf{S} \mathbf{V}^T
\end{equation}

where $\mathbf{U} \in \mathbb{R}^{n \times n}$ and $\mathbf{V} \in \mathbb{R}^{p \times n}$ contain the left and right singular vectors and where $\mathbf{S} \in \mathbb{R}^{p \times p}$ is the diagonal matrix of singular values. Importantly, $\mathbf{U}$ and $\mathbf{V}$ have rank equal to that of $\mathbf{X}$. The $i$th principal component, then, is the $i$th column of $\mathbf{U}$. The corresponding column of $\mathbf{V}$ gives weights that indicate the extent to which each PANAS-X category contributed to that component. Similarly, squaring the corresponding singular element of $\mathbf{S}$ gives the magnitude of variance accounted for by that component. In the \textbf{Supplemental Information} we examine, in detail, the robustness of this analysis to the exclusion of individual data points (Figure~\ref{figure:leaveOneOut}), random permutations of the flexibility estimates (Figure~\ref{figure:randomMood}), eschewing PCA in favor of individual PANAS-X categories (Figure~\ref{figure:individualPANASXCategories}), other principal components besides the first (Figure~\ref{figure:PC4}), the use previously-defined emotional affect classes rather than a principal component to identify indices of positivity (Figure~\ref{figure:panasClasses}--\ref{figure:fatigue}), and the use of exploratory factor analysis instead of PCA (Figure~\ref{figure:figEFA}). We also test the robustness of our results after controlling for other psychophysiological and nuisance variables (Figure~\ref{figure:headMotion},\ref{figure:nuisance}).

\subsection*{Dynamic network construction and community detection}
We sought a division of brain regions into communities, which are thought to reflect the brain's functional sub-systems \cite{sporns2016modular}. We divided the parcel time series into $T=14$ windows of 37 time points (TRs) each ($\approx$1.36 minutes in length). For each window, we calculated the wavelet coherence matrix, $\mathbf{C} \in \mathbb{R}^{N\times N}$. Each element, $C_{ij}$, represented the magnitude squared coherence of the scale two (0.0625--0.125 Hz) Daubechies wavelet (length 4) decomposition of the windowed time series obtained from regions $i$ and $j$ (\url{http://www.atmos.washington.edu/~wmtsa/}). This particular frequency band was selected based on previous work \cite{bassett2011dynamic, braun2015dynamic}. Each dynamic network was treated as a layer in a multi-layer network, $\mathcal{C} = \{ \mathbf{C}_1 \ldots \mathbf{C}_{14} \}$. To detect the temporal evolution of modules, we maximized the multi-layer modularity \cite{mucha2010community}, which seeks the assignment all brain regions in all layers to modules such that:

\begin{equation}
\begin{split}
Q(\gamma, \omega) & = \frac{1}{2 \mu} \sum_{ijsr} [(C_{ijs} - \gamma P_{ijs})]\delta (G_{is},G_{js}) \\
& + \delta (i,j) \cdot \omega ]\delta (G_{is}, G_{jr})
\end{split}
\end{equation}

is maximized. In this expression, $C_{ijs}$ is the coherence of regions $i$ and $j$ in layer $s$. The tensor $P_{ijs}$ is the expected coherence in an appropriate null model. Specifically, we choose $P_{ijs} = \frac{k_{is} k_{js}}{2m_s}$, which is a multi-layer extension of the common configuration model. The parameter, $\gamma$, scales the relative contribution of the expected connectivity and effectively controls the number of modules detected within a given layer. The other free parameter, $\omega$, determines the similarity of modules across layers, and is therefore sometimes referred to as the \emph{temporal resolution parameter} \cite{bassett2013robust}. In the main text, we fix these parameters to the commonly-used default values of $\gamma = \omega = 1$ \cite{bassett2013robust}. In the supplement we demonstrate the robustness of our results to variation in the resolution parameters (Figure \ref{figure:paramSweep}).

We use a Louvain-like locally greedy algorithm to maximize the multi-layer modularity, $Q(\gamma, \omega)$ \cite{jutla2011generalized} (typical output is show in Figure~\ref{figure:exampleLouvainOutput}). Due to near-degeneracies in the modularity landscape \cite{good2010performance} and stochastic elements in the optimization algorithm \cite{blondel2008fast}, the output typically varies from one run to another. For this reason, rather than focus on any single run, we characterized the statistical properties of 50 runs of the algorithm, which correspond to 50 optimizations of the multi-layer modularity.

\subsection*{Regional and global flexibility}
The output of the Louvain-like locally greedy algorithm was a partition, $\mathbf{G} \in \mathbb{R}^{N \times T}$, whose element $G_{i,r} \in \{1 \ldots c \}$ is the community to which brain region $i$ in layer $r$ was assigned in that optimization. The multi-layer modularity maximization simultaneously assigns brain regions in all layers to communities so that community labels are consistent across layers, thus circumventing the commonly studied community matching problem. Given $\mathbf{G}$, we can calculate each brain region's flexibility score:

\begin{equation}
f_i = \frac{1}{T - 1} \sum_{s = 1}^{T - 1} \delta (G_{i,s},G_{i,s + 1})
\end{equation}

which counts the fraction of times that brain region, $i$, changes its community assignment in successive layers. Flexibility is normalized so that scores near zero and one correspond to brain regions whose community assignments are highly consistent and highly variable, respectively, across layers. Flexibility can also be averaged over all brain regions to obtain the \textit{global flexibility} of the whole brain, $F = \frac{1}{N} \sum_{i = 1}^N f_i$. Both regional and global flexibility scores were calculated separately for each of the 50 modular partitions obtained from the Louvain-like algorithm and averaged across optimizations.

\subsection*{System assignments}
In addition to community assignments, each brain region was also assigned to a brain system. The designation of regions to systems was developed as part of an earlier study \cite{laumann2015functional}, which defined, in total, 13 systems based on the topographic patterns of functional connections: cingulo-opercular (CO), default mode (DMN), dorsal attention (DAN), fronto-parietal 1 (FP1), fronto-parietal 2 (FP2), medial-parietal (MedPar), parietal-occipital (ParOcc), salience (SAL), somatomotor (SMN), ventral attention (VAN), primary visual (VIS1), and peripheral visual (VIS2). In addition to the previously-defined systems \cite{laumann2015functional} we also included two other systems: (i) a subcortical (SUB) system comprised of bilateral thalamus, caudate, putamen, pallidum, hippocampus, amygdala, and accumbens, and (ii) a category reserved for brain regions with no clear assignment (NONE) (See Figure~\ref{figure:brainSystems}). 

In the main text, it was sometimes advantageous to describe measures at the level of brain systems as opposed individual brain regions or the whole brain (e.g. mean and standard deviation flexibility; mean correlation of flexibility with the positivity index). Such system-level measures were obtained by averaging across each system's constituent regions. However, because such measures may be biased by system size (i.e. number of regions assigned to that system), we compared the observed measures against the distribution of similar measurements obtained from a permutation null model, wherein the total number of regions assigned to each system remained constant but where assignments were, otherwise, made at random. Specifically, we calculated the mean, $\mu_{sys}$, and standard deviation, $\sigma_{sys}$, system-level measurements based on 10000 iterations of the permutation null model and expressed the observed measure, $y_{sys}$, as a $z$-score:

\begin{equation}
z_{sys} = \frac{y_{sys} - \mu_{sys}}{\sigma_{sys}}\\.
\end{equation}

We corrected for multiple comparisons by controlling the false discovery rate (FDR) using the linear step-up procedure \cite{benjamini1995controlling}. In each case, we calculated an adjusted critical value, $p_{adj}$, by fixing the maximum FDR at $d=0.001$.

\subsection*{Acknowledgements} We thank Tyler Moore for helpful discussions and Russell Poldrack for providing additional imaging resources. RFB and DSB acknowledge support from the John D. and Catherine T. MacArthur Foundation, the Alfred P. Sloan Foundation, the Army Research Laboratory and the Army Research Office through contract numbers W911NF-10-2-0022 and W911NF-14-1-0679, the National Institute of Mental Health (2-R01-DC-009209-11), the National Institute of Child Health and Human Development (1R01HD086888-01), the Office of Naval Research, and the National Science Foundation (BCS-1441502 and BCS-1430087). JIG acknowledges support from the National Science Foundation (NSF-1533623. TDS was supported by the National Institute of Mental Health (K23MH098130 and R01MH107703).

\clearpage
\beginsupplement
\section*{Supplementary Information}
\subsection{Community detection}
\subsubsection{Multi-layer modularity maximization}

A central topic in network science, generally, is how to identify a network's mesoscale structure: what are the organizational principles that dominate the intermediate scale between that of individual nodes and the network as a whole \cite{newman2012communities}? One possible class of mesoscale organization is ``community structure,'' meaning that the network can be decomposed into communities or modules of densely inter-connected sub-networks \cite{fortunato2010community}. The most popular method for detecting communities is to partition a network's nodes into non-overlapping clusters so as to maximize a modularity quality function of the form \cite{newman2004finding}: $Q = \text{observed within-community density} - \text{expected within-community density}$. The partition that maximizes $Q$ is often accepted as a good estimate of a network's community structure.

More formally, the typical expression for $Q$ is given as:

\begin{equation}
Q = \frac{1}{2m} \sum_{ij}^N [C_{ij} - P_{ij}] \delta (G_i,G_j)\\.
\end{equation}

where, $C_{ij}$ and $P_{ij}$ are the observed and expected weight of the connection between nodes $i,j$, respectively. Often, the expected weight is given as: $P_{ij} = \frac{k_i k_j}{2m}$, which corresponds to the null model wherein connections are formed randomly but where node strengths (i.e. $k_i = \sum_j C_{ij}$) are preserved. The Kronecker delta function is equal to 1 when nodes' community assignments are the same (i.e. $G_i = G_j$) and is 0 otherwise, ensuring that the only $i,j$ contibuting to the summation (and therefore to $Q$) are those assigned to the same community.

Recently, modularity has been generalized so as to be compatible with multi-layer networks \cite{mucha2010community}. A multi-layer network extends the standard network model of nodes and edges to one in which additional layers can be used to represent different connection classes among nodes (e.g. air, road, rail traffic between cities) or, in our case, observations of the same network at different time points \cite{kivela2014multilayer}. The expression for modularity's multi-layer analogue retains the general form of the single-layer version:

\begin{equation}
\begin{split}
Q(\gamma, \omega) & = \frac{1}{2 \mu} \sum_{ijsr} [(C_{ijs} - \gamma P_{ijs})]\delta (G_{is},G_{js}) \\
& + \delta (i,j) \cdot \omega ]\delta (G_{is}, G_{jr})
\end{split}
\end{equation}

\begin{figure}[b]
\begin{center}
\centerline{\includegraphics[width=0.5\textwidth]{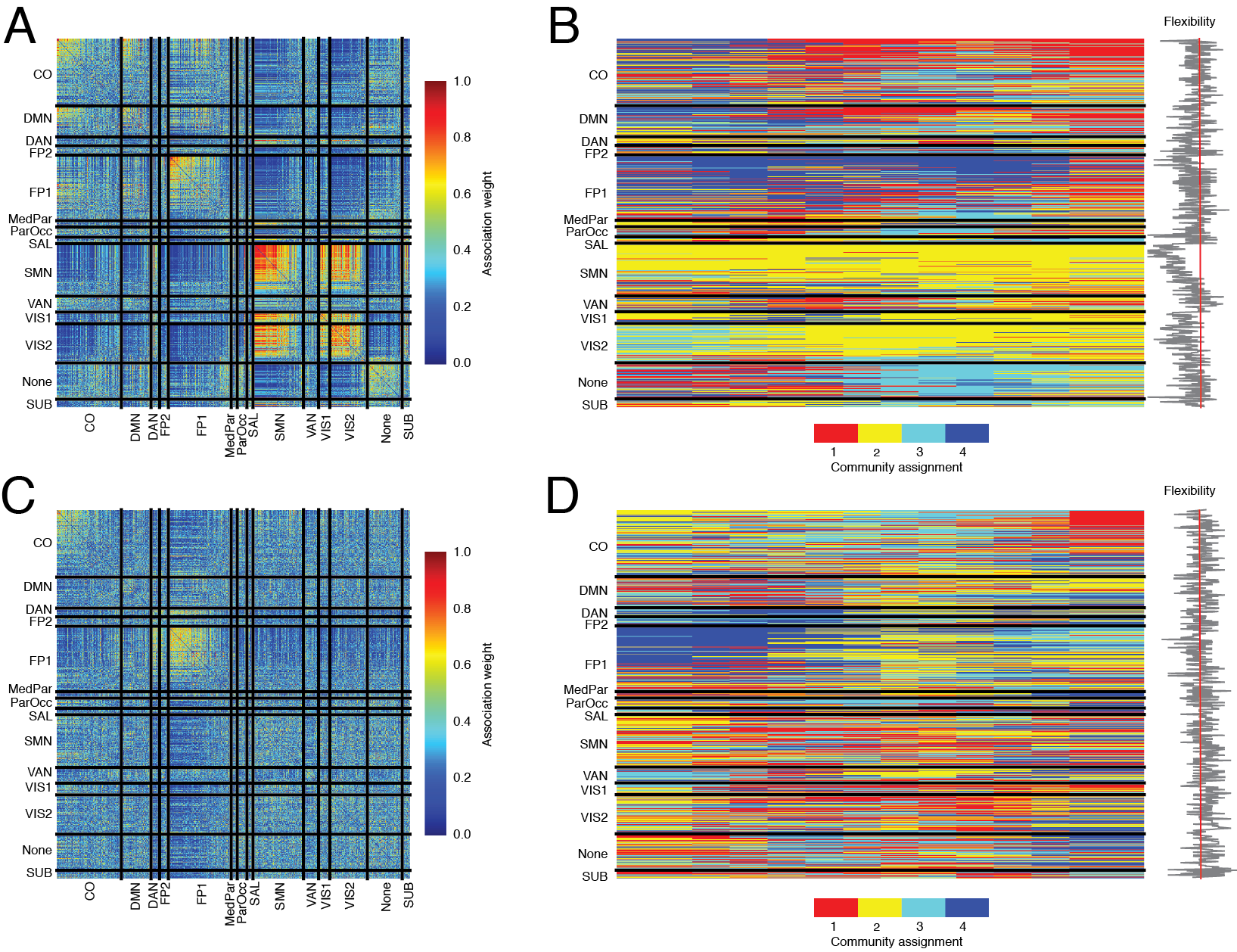}}
\caption{\textbf{Example ``low'' and ``high'' flexibility sessions.} \emph{(A,B)} Association matrix for low \emph{versus} high flexibility session. \emph{(C,D)} Dynamic community matrix and regional flexibility series.}\label{figure:exampleLouvainOutput}
\end{center}
\end{figure}

Here, $C_{ijs}$ is the weight of the connection between nodes $i,j$ in layer $s$ and $P_{ijs}$ is the corresponding expected weight in an appropriate null model. The variable, $G_{is}$, encodes the community assignments of node $i$ in layer $s$. Typical output of the algorithm is shown in Figure \ref{figure:exampleLouvainOutput}B,D.

Modularity, both single- and multi-layer, can be maximized using a number of algorithms \cite{fortunato2010community}. Perhaps the most popular is the so-called ``Louvain algorithm'' \cite{blondel2008fast}. Due to the stochastic nature of the algorithm, the computational complexity of modularity maximization, and the fact that the modularity landscape usually features many near-optimal solutions \cite{good2010performance}, it is considered best practice not to focus on the output of any single run, but to run the algorithm many times and characterize the statistical properties of an ensemble of near-optimal solutions \cite{bassett2013robust}.

\subsubsection{Robustness to choice of resolution parameters}
The expression for multi-layer modularity further generalizes modularity by introducing two resolution parameters, $\gamma$ and $\omega$. Along with the actual structure of the network itself, these parameters determine the composition of detected communities. Specifically, $\gamma$ scales the importance of the expected weight (the term $P_{ijs}$), effectively determining community number and size. The parameter $\omega$, on the other hand, controls the strength of inter-layer coupling between nodes; larger or smaller values of $\omega$ lead to more- or less-consistent communities across layers.

\begin{figure}[t]
\begin{center}
\centerline{\includegraphics[width=0.5\textwidth]{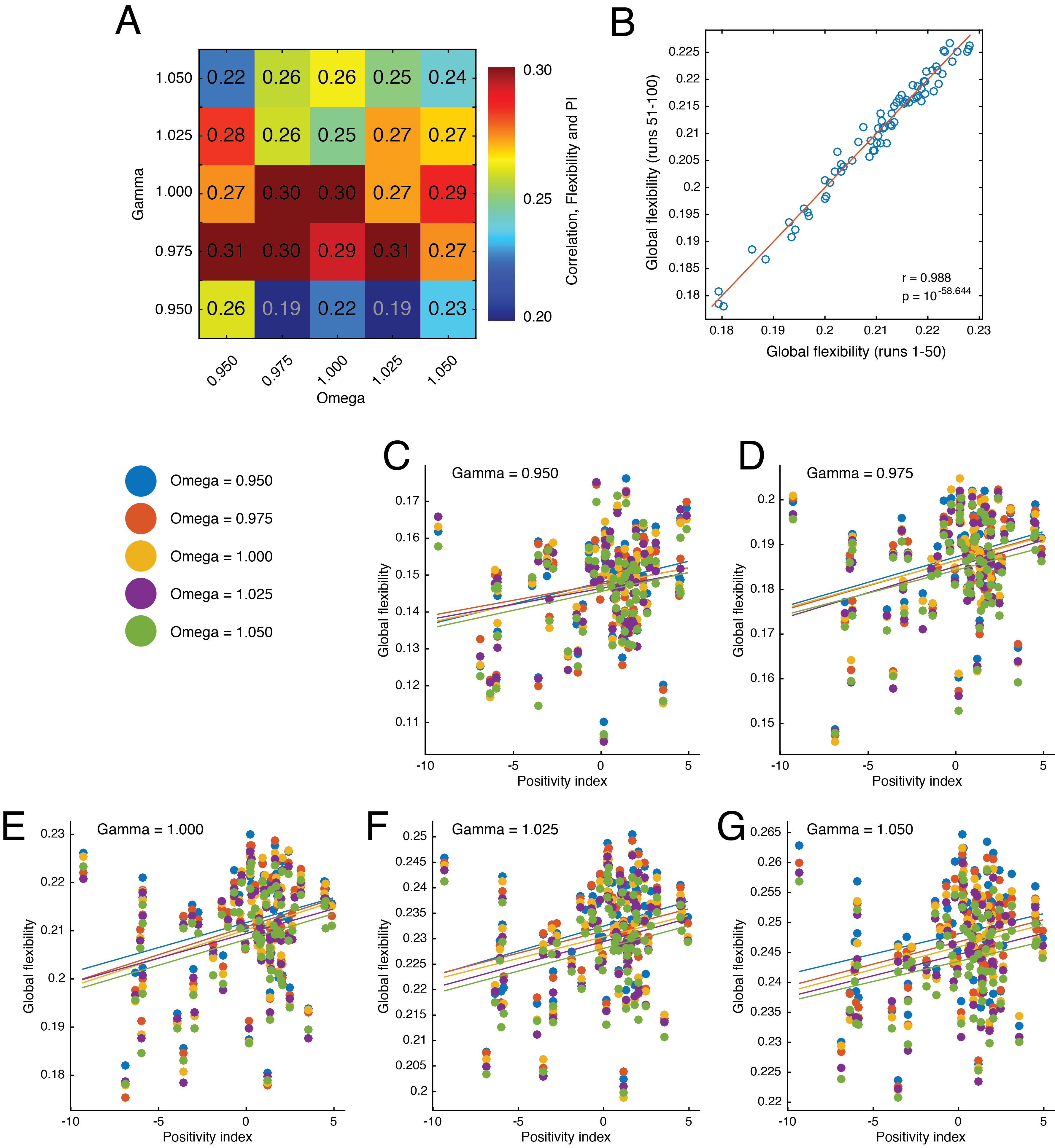}}
\caption{\textbf{Robustness to choice of resolution parameters.} \emph{(A)} Pearson correlation coefficients for 25 different $\gamma,\omega$ values in the neighborhood of $\gamma = 1, \omega = 1$. \emph{(B)} Scatterplot of global flexibility scores obtained from Louvain runs 1-50 and 51-100. \emph{(C-G)} Scatterplots of positivity index, $PI$, against global flexibility for each of the 25 parameter combinations. Each panel fixes $\gamma$ at a particular value (either 0.950, 0.975, 1.000, 1.025, 1.050) and varies $\omega$ over the same range.}\label{figure:paramSweep}
\end{center}
\end{figure}

Despite the importance of these parameters, many applications of multi-layer modularity leave them fixed at their \textit{de facto} default values, $\gamma = \omega = 1$ \cite{bassett2013robust}. Though a complete exploration of both parameters is beyond the scope of most studies, it is considered good practice to demonstrate that one's results are robust to reasonable variations in the parameters' values \cite{bassett2011dynamic,bassett2013robust,braun2015dynamic}. Accordingly, we sought to replicate the principal results from the main text, i.e. the correlation of flexibility with positivity, across variations in the two resolution parameters. To this end, we varied both $\gamma$ and $\omega$ over the range [0.95, 0.975, 1.00, 1.025, 1.05] and maximized multi-layer modularity for all pairs of parameter values. This procedure generated 25 estimates of $\hat{r}(PI,F)$ (24 \textit{new} estimates, including a repeat of the case where $\gamma = \omega = 1$, which was already investigated in the main text). In general, these results support the hypothesis that $\hat{r}(PI,F)>0$ (minimum correlation of $\hat{r}(PI,F) = 0.19$ and maximum correlation of $\hat{r}(PI,F) = 0.31$) (Figure \ref{figure:paramSweep}A). We also took this opportunity to cross-validate the flexibility estimates we obtained in the main text by comparing those estimates to the new set of flexibility estimates obtained with $\gamma = \omega = 1$. We found remarkable consistency, with $\hat{r} = 0.988$ and $p < 10^{-15}$ (Figure \ref{figure:paramSweep}B). For completeness, we show scatterplots of flexibility against the positivity index for all $\gamma, \omega$ pairs (Figures \ref{figure:paramSweep}C-G).

\subsection{Robustness of positivity index with global flexibility correlation}

In the main text, we presented data suggesting that an index of positivity, $PI$, was positively correlated with global flexibility, $F$, i.e. $\hat{r}(PI,F) > 0$. This main result indicates that more flexible brains, i.e. greater dynamic reconfiguration, are associated with positive emotions. However, it is important to acknowledge that the Pearson's correlation coefficient, which we used to characterize the relationship between these variables, can be sensitive to outlying data points, which can bias the overall magnitude of the observed correlation coefficient. To quantify the robustness of our reported estimates, we assessed the robustness of the observed correlation coefficient value using a ``leave one out'' analysis. In brief, this analysis consists of performing the principal component analysis (PCA) using, instead of the full set of PANAS-X variables, a limited set, thereby yielding new estimates of the positivity index, which we call $PI'$. Then, we calculate the correlation between the positivity index and flexibility $\hat{r}(PI',F)$ and determine whether the results of this analysis are in line with those presented in the main text.

\begin{figure}[b]
\begin{center}
\centerline{\includegraphics[width=0.5\textwidth]{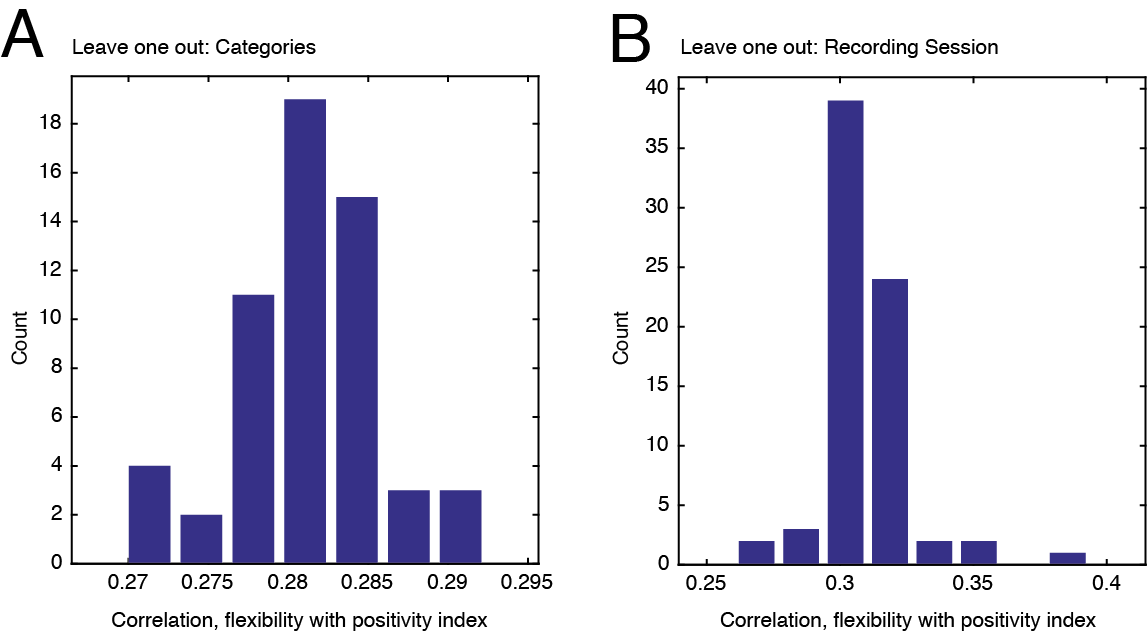}}
\caption{\textbf{Summary of robustness testing procedure.} \emph{(A)} Distribution of correlation coefficients, $\hat{r}(PI',F)$, where $PI'$ is a positivity index estimated after omitting PANAS-X categories from the PCA analysis. \emph{(B)} Distribution of correlation coefficients, $\hat{r}(PI',F)$, where $PI'$ is a positivity index estimated after omitting all data acquired from each recording session from the PCA analysis.}\label{figure:leaveOneOut}
\end{center}
\end{figure}

We generate limited sets of observations two different ways, each corresponding to a distinct null hypothesis. First, to test the hypothesis that the correlation between the positivity index and flexibility $\hat{r}(PI,F)$ is biased by individual PANAS-X categories, we systematically omit each of the $p=57$ categories from the data set, performing the PCA using all $n=73$ observations but only $p-1=56$ categories. Second, to test the hypothesis that $\hat{r}(PI,F)$ is biased by individual recording sessions, we systemically omit each of the $n = 73$ recording sessions, which includes all the PANAS-X categories as well as the global flexibility score. For this second test, we perform PCA on $n - 1=72$ observations of $p$ categories, and calculate $\hat{r}(PI',F)$ based on the same $n-1=72$ observations.

\begin{figure}[t]
\begin{center}
\centerline{\includegraphics[width=0.5\textwidth]{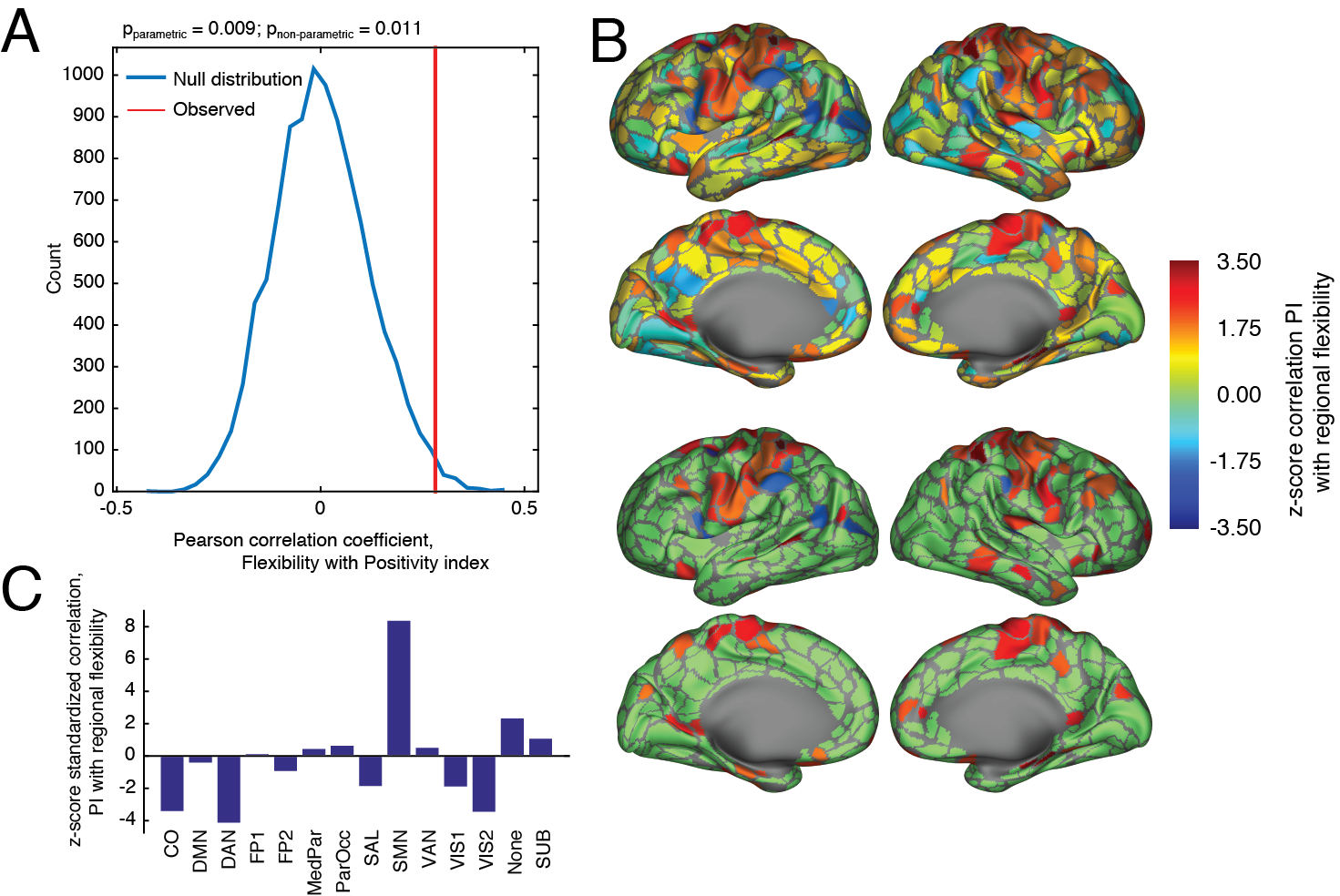}}
\caption{\textbf{Summary of randomly permuted recording sessions.} \emph{(A)} Distribution of correlation coefficients of $PI$ with $F$ after randomly permuting the order of the recording sessions. The red line indicates the observed correlation coefficient. \emph{(B)} Topographic representation of z-score correlation of regional flexibility with $PI$ obtained after randomly permuting the order recording sessions. The top panel shows the complete set of z-scored coefficients; the bottom panel shows only those regions whose z-score exceed $\pm2$. \emph{(C)} Standardized z-score correlation of regional flexibility with $PI$ aggregated by system.}\label{figure:randomMood}
\end{center}
\end{figure}

In both cases, we found ample support for the hypothesis that $\hat{r}(PI,F) > 0$. Moreover, for every $PI'$ we generated, the correlation magnitude always satisfied $0.26 \ge \hat{r}(PI',F) \ge 0.39$ (Figure \ref{figure:leaveOneOut}A,B). While not conclusive, these analyses provided additional evidence that emotional state is positively correlated with global flexibility and not obviously biased by any single observation.

In addition to these ``leave one out'' analysis, we also tested the null hypothesis that the magnitude of the observed correlation $\hat{r}(PI,F)$ occurred by chance. To this end, we randomly permuted the order of the vector containing the global flexibility scores and calculated the correlation of the re-ordered flexibility with $PI$. We repeated this 10000 times, generating a null distribution of correlation coefficients, against which we compared the observed correlation. We performed this comparison both parametrically and non-parametrically. The parametric test involved estimating the mean, $\hat{r}_{\mu}(PI,F)$, and standard deviation, $\hat{r}_{\sigma}(PI,F)$, of the null distribution and standardizing the observed correlation against these estimates: $\hat{r}_z(PI,F) = \frac{\hat{r}(PI,F) - \hat{r}_{\mu}(PI,F)}{\hat{r}_{\sigma}(PI,F)}$. This non-parametric test involved, simply, estimating what fraction of null distribution was greater than the observed correlation coefficient. The parametric and non-parametric tests yielded p-values of $p=0.0087$ and $p=0.0111$, respectively. These results suggest that the observed correlation, $\hat{r}(PI,F)$, was not likely to have been produced by chance (Figure \ref{figure:randomMood}A).

We performed an analogous analysis of the regional flexibility scores, which generated region-level flexibility z-scores (Figure \ref{figure:randomMood}B). As in the main text, we aggregated these scores by brain system and, once again, compared the average z-score for each system against a permutation-based chance model. The system with the single greatest z-score was the somatsensory system ($z_{SMN} = 8.369$, $p < 10^{-15}$), which agrees with the results presented in the main text (Figure \ref{figure:randomMood}C).

\subsection{Correlation of PANAS-X categories with global flexibility}

\begin{figure}[b]
\begin{center}
\centerline{\includegraphics[width=0.5\textwidth]{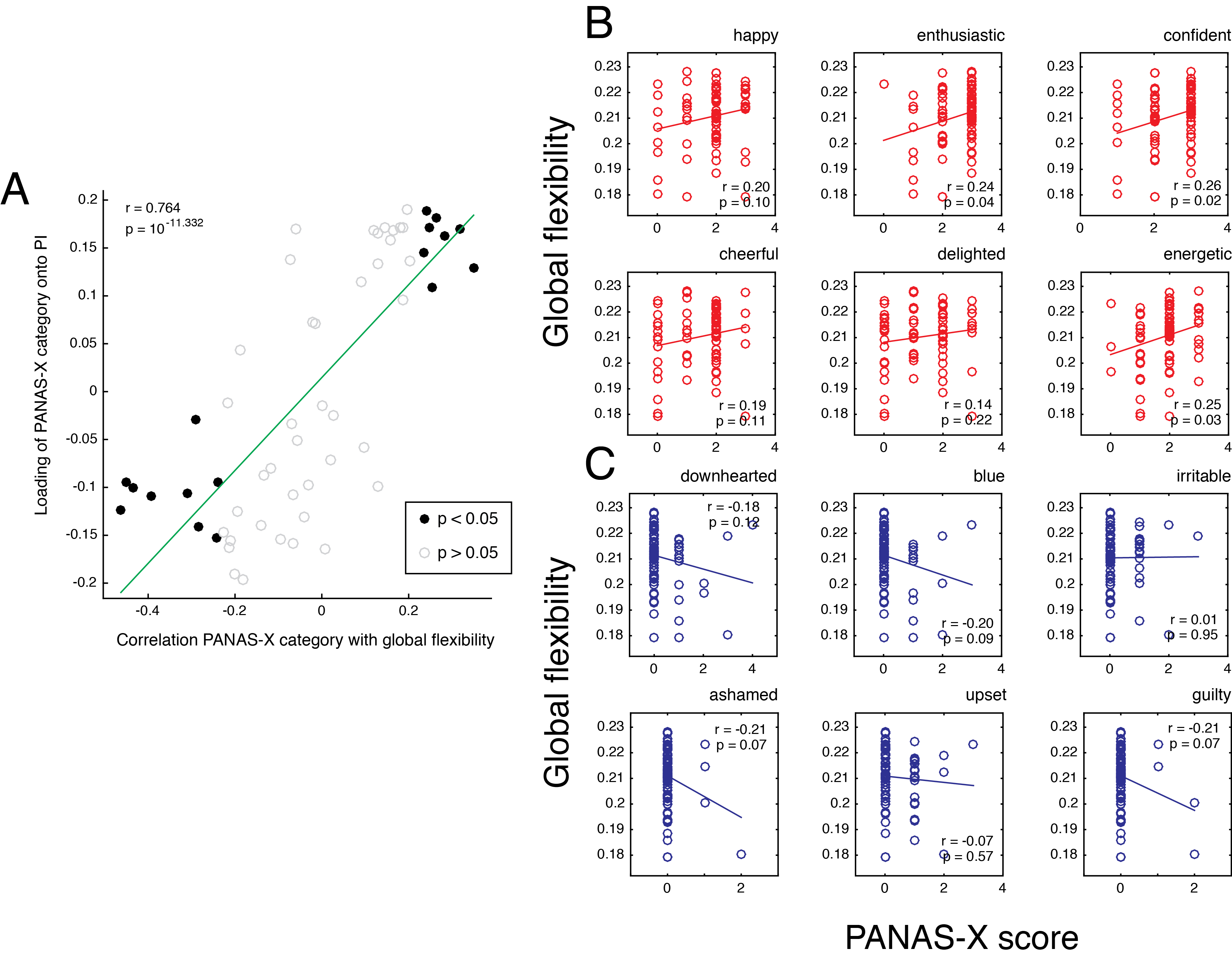}}
\caption{\textbf{Relationship between flexibility and individual PANAS-X scores.} \emph{(A)} Relationship of individual PANAS-X categories to PANAS-X loadings onto $PI$. \emph{(B)} Top six PANAS-X categories, in terms of loading onto $PI$, and their relationships with global flexibility, $F$. \emph{(C)} Bottom six PANAS-X categories, in terms of loading onto $PI$, and their relationships with global flexibility, $F$.}\label{figure:individualPANASXCategories}
\end{center}
\end{figure}

The results presented in the main text showed a positive correlation of global flexibility with a ``positivity index'' ($PI$). The positivity index itself was based on a PCA of PANAS-X categories. Though the loadings of those categories onto $PI$ gives some sense of how individual emotional categories contribute to $PI$, we nonetheless felt that it would be illuminating to present the reader with plots showing the relationship of PANAS-X categories with flexibility. Indeed, the correlation magnitude of PANAS-X categories with $PI$ was highly correlated with the loadings of each category (Figure \ref{figure:individualPANASXCategories}A). Figures \ref{figure:individualPANASXCategories}B,C show, in detail, the PANAS-X categories with the strongest positive loadings onto $PI$ (shown in red in Figure \ref{figure:individualPANASXCategories}B) and those with the strongest negative loadings onto $PI$ (shown in blue Figure \ref{figure:individualPANASXCategories}C).

\begin{figure}[t]
\begin{center}
\centerline{\includegraphics[width=0.5\textwidth]{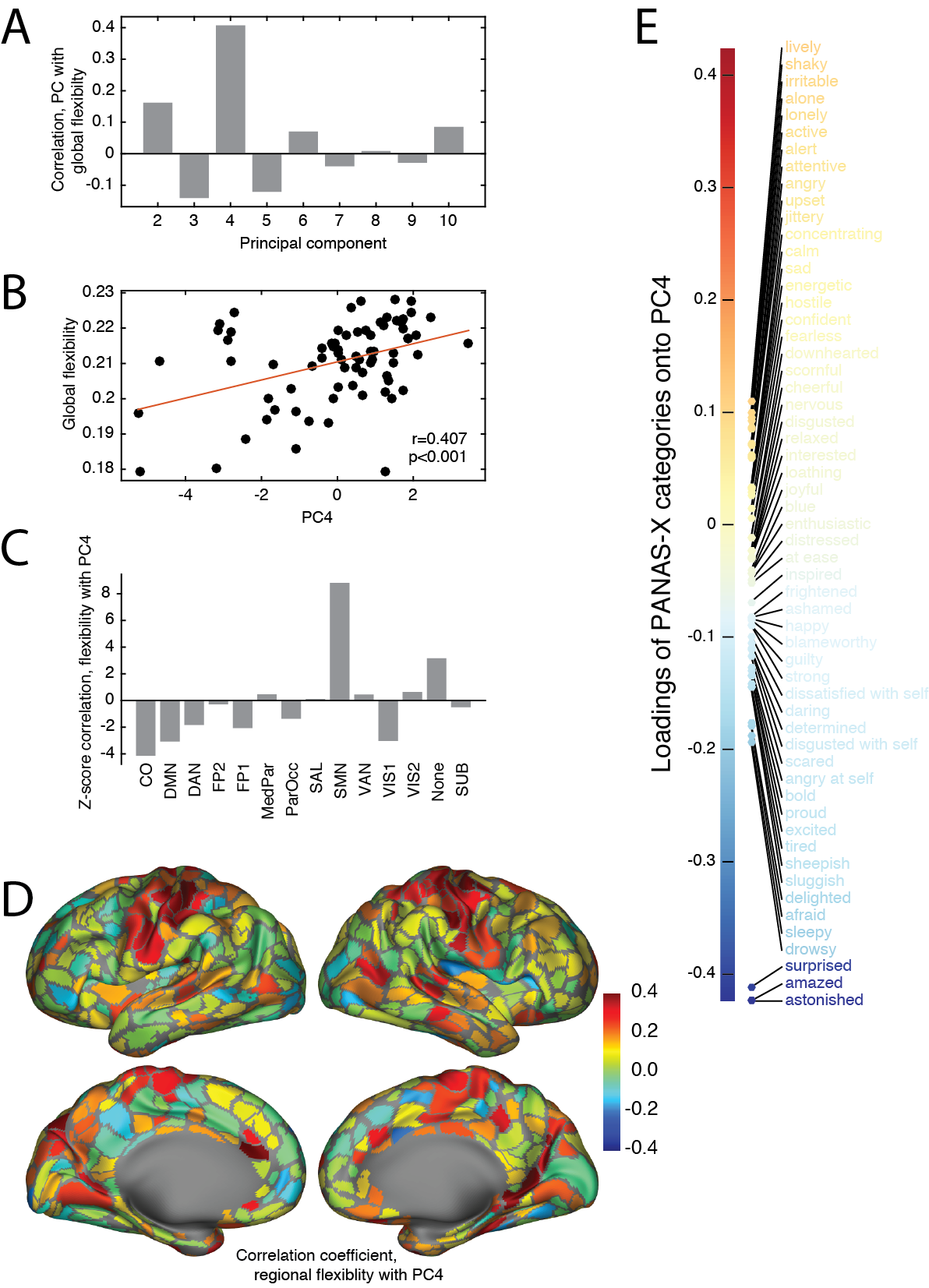}}
\caption{\textbf{Analysis of other principal components.} \emph{(A)} Correlation magnitude for the first ten principal components. \emph{(B)} Scatterplot of global flexibility against the fourth component, $PC4$. \emph{(C)} Correlation of regional flexibility scores with $PC4$ and averaged by brain system. The $y$-axis is expressed as $z$-scores relative to a permutation null model. \emph{(D)} Topographic representation of the correlation magnitude of regional flexibility with $PC4$. \emph{(E)} Loadings of PANAS-X scores onto $PC4$.}\label{figure:PC4}
\end{center}
\end{figure}

\subsection{Other principal components}
We termed the first principal component the ``positivity index'' ($PI$), which accounted for an overwhelmingly large percentage of variance ($\approx 33\%$) among the PANAS-X scores. Despite the other principal components accounting for much less total variance, we sought to investigate whether any were predictive of global flexibility, $F$. To this end, we calculated the correlation between flexibility and the first ten principal components ($PC2, \ldots, PC10$), excluding $PC1$, which we already analyzed as $PI$. These remaining components accounted for, at most $\approx 11\%$ and no less than $\approx 2\%$, of the remaining variance. The fourth component, $PC4$,  accounted for $\approx 6\%$ variance, and was correlated with, $F$ ($\hat{r}(PC4,F) = 0.407$, $p = 3.49 \times 10^{-4}$) (Figure \ref{figure:PC4}A,B). The remaining components were not obviously correlated with global flexibility (mean correlation magnitude of $3.53\times10^{-4}$, after excluding $PC4$). Unlike $PI$, which placed PANAS-X categories along a continuum of ``positive'' and ``negative'' emotions, $PC4$ paints a less clear picture, intermingling PANAS-X terms of seemingly dissimilar emotional affect (e.g. ``proud'' and ``disgusted with self'' have similar loadings). Nonetheless, we can speculate that $PC4$ tracks the subject's state of arousal or surprise based on the observation that the PANAS-X categories with the greatest magnitude loadings were ``surprised'', ``amazed'', and ``astonished'' (Figure \ref{figure:PC4}E). Finally, we calculated the correlation coefficients of regional flexibility scores with $PC4$ and aggregated them by brain systems. The most obvious association was the positive correlation of regional flexibility and $PC4$ for regions within somatomotor cortex (permutation test; $z_{SMN} = 8.81$, $p < 10^{-15}$) (Figure \ref{figure:PC4}C,D). Collectively, these results suggest that increased levels of surprise correspond to decreases in global flexibility. Furthermore, this relationship appears to have its neuroanatomical underpinnings in somatomotor cortex.

\begin{figure}[b]
\begin{center}
\centerline{\includegraphics[width=0.5\textwidth]{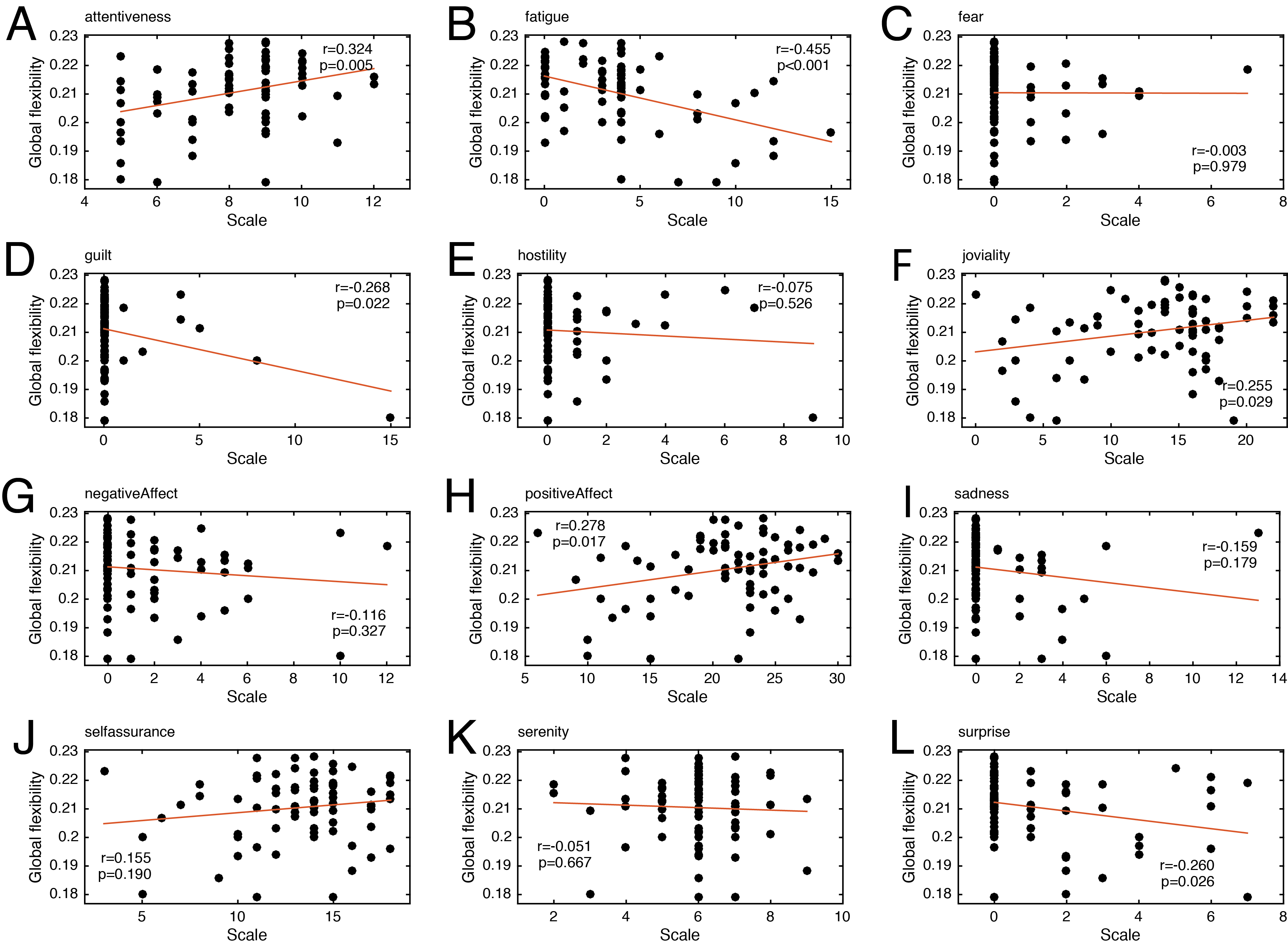}}
\caption{\textbf{Correlations of flexibility with PANAS-X affect classes.} \emph{(A-L)} Scatterplot of PANAS-X composite affect scores with global flexibility.}\label{figure:panasClasses}
\end{center}
\end{figure}

\begin{figure}[t]
\begin{center}
\centerline{\includegraphics[width=0.5\textwidth]{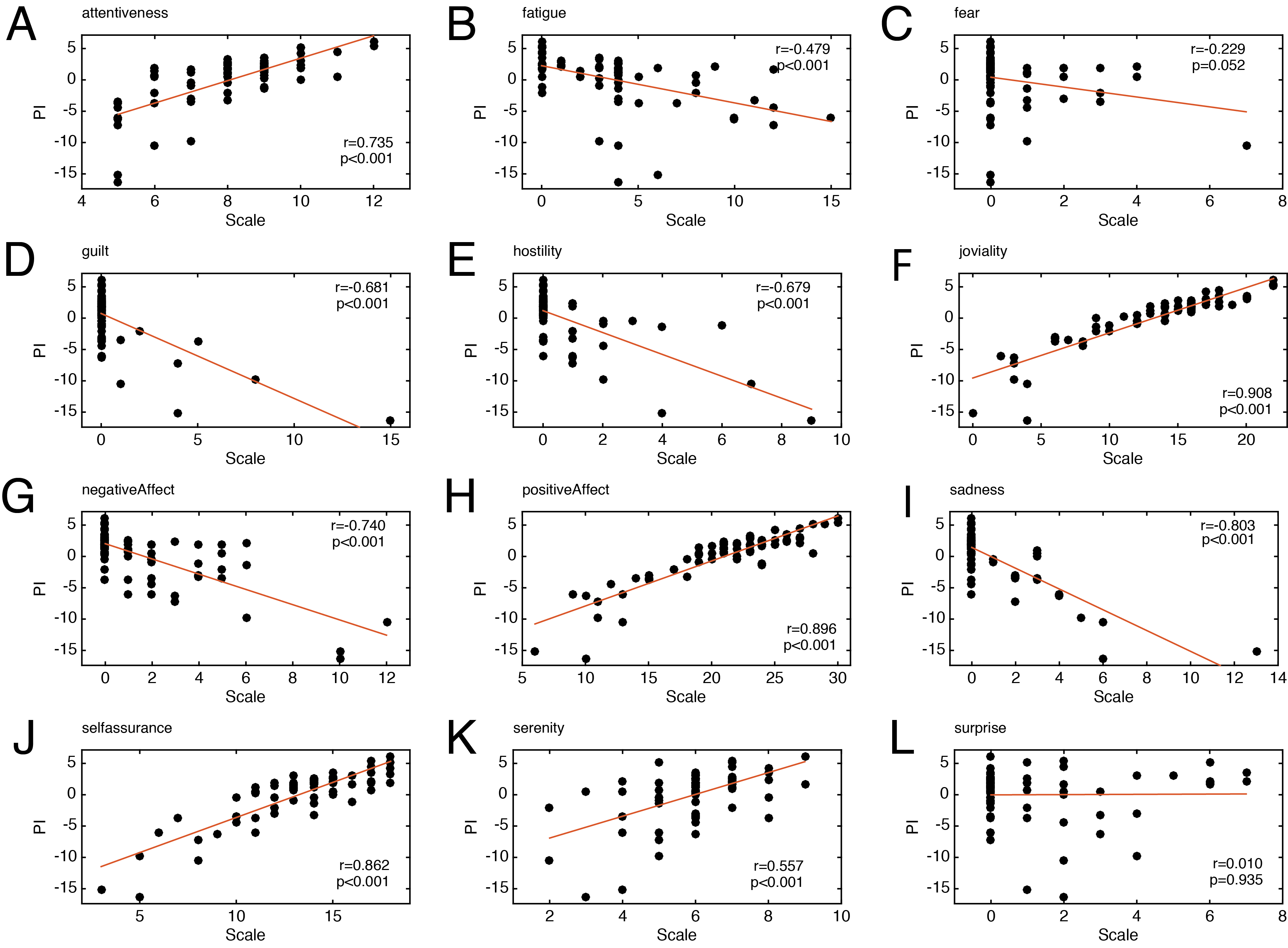}}
\caption{\textbf{Correlations of positivity index with PANAS-X affect classes.} \emph{(A-L)} Scatterplot of PANAS-X composite affect scores with positivity index.}\label{figure:PANASClassesVsPI}
\end{center}
\end{figure}

\begin{figure}[b]
\begin{center}
\centerline{\includegraphics[width=0.5\textwidth]{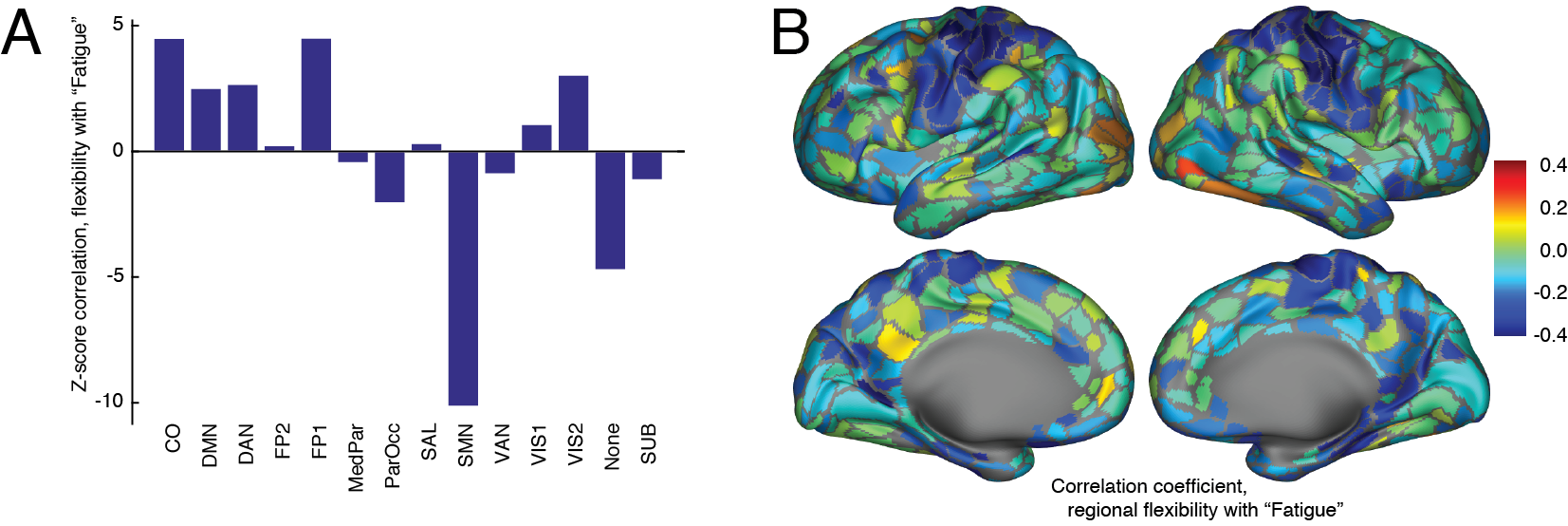}}
\caption{\textbf{Correlations of regional flexibility with ``fatigue''.} \emph{(A)} System-level correlations. \emph{(B)} Topographic distribution showing the correlation magnitude of ``fatigue'' with regional flexibility.}\label{figure:fatigue}
\end{center}
\end{figure}

\subsection{PANAS-X affect scores}
The PANAS-X features (in the form used here) $p=57$ categories. In the main text we used a PCA to distill these categories into a single composite ``positivity index''. Nonetheless, there exist alternative methods for grouping the categories into other affective classes. One commonly used set of sub-divisions \cite{watson1999panas} groups categories into hierarchical affect classes: ``negative affect'', ``positive affect'', ``fear'', ``hostility'', ``guilt'', ``sadness'', ``joviality'', ``self-assurance'', ``attentiveness'', ``fatigue'', ``serenity'', and ``surprise'' (Table \ref{table:panasxCategories}). Often ``shyness'' is also included, though the categories comprising this class were omitted from the PANAS-X used in the current study. A composite score can be estimate for each affect class by summing the responses of the PANAS-X categories. We tested whether global flexibility was correlated with any of these affect classes. Indeed, a number of these classes exhibited significant correlations with flexibility. At a statistical threshold of $p < 0.05$ (uncorrected), we observed that ``positive affect'' ($\hat{r} = 0.278, p = 0.017$), ``guilt'' ($\hat{r} = -0.268, p = 0.022$), ``joviality'' ($\hat{r} = 0.255, p = 0.029$), ``attentiveness'' ($\hat{r} = 0.324, p = 0.051$), ``surprise'' ($\hat{r} = -0.260, p = 0.026$), and ``fatigue'' ($\hat{r} = -0.455, p = 0.001$) were all correlated with flexibility (Figure \ref{figure:panasClasses}). In general, these results agree with those obtained from our PCA analysis: ``positive affect'', ``joviality'', and ``attentiveness'' (which capture positive emotions, broadly) are all positively correlated with flexibility and also strongly correlated with the PCA-derived positivity index ($\hat{r} = 0.896, 0.908, 0.735$, respectively; all $p < 10^{-15}$). Less positive terms like ``guilt'', ``fatigue'', and ``surprise'', on the other hand, were all negatively correlated with flexibility and either negatively correlated or uncorrelated with the positivity index ($\hat{r} = -0.681, -0.479, 0.010$, $p = 3.341 \times 10^{-11}, 1.820 \times 10^{-5}, 0.935$, respectively) (Figure \ref{figure:PANASClassesVsPI}). Interestingly, the PANAS-X categories that comprise the ``surprise'' class were identical to those with the strongest loadings onto the fourth principal component, $PC4$. 

The affect class ``fatigue'' exhibited the greatest magnitude correlation with global flexibility. Accordingly, we investigated its correlation with regional flexibility scores to determine its topographic distribution and its mapping onto brain systems. Like the positivity index, $PI$, and other principal components, e.g. $PC4$, ``fatigue'' was strongly anti-correlated with the regional flexibility of somatomotor cortex ($z_{SMN} = -10.110, p_{SMN} < 10^{-15}$) (Figure \ref{figure:fatigue}A,B). Also as before, a number of other systems are also implicated, including cingulo-opercular (CO), fronto-parietal (FP1), and peripheral visual (VIS2) ($z_{CO} = 4.469$, $z_{FP1} = 4.481$, $z_{VIS2} = 3.002$; all $p < 0.001$), which were more positively correlated with fatigue than expected by chance.

\subsection{PCA versus factor analysis}
PCA and exploratory factor analysis (EFA) both attempt to identify low-rank approximations of the covariance structure among a set of observed variables. Both techniques have been applied widely in the psychological sciences, where they have been instrumental in the development of clinical and behavioral indices and the subsequent identification of items that load onto these scales \cite{comrey1988factor, galbraith2002analysis}. The rationale for using either technique in place of the other is part of an ongoing debate \cite{fabrigar1999evaluating}. In this subsection, we demonstrate that under some reasonable assumptions, EFA yields similar results to those presented in the main text for PCA.

\subsubsection{PCA model}
Essentially, PCA assumes that a matrix of observations, $\mathbf{X} \in \mathbb{R}^{n \times p}$, can be decomposed into principal components, where each component has the form:

\begin{equation}
\mathbf{c}_i = w_{1i} \mathbf{x}_1 + w_{2i} \mathbf{x}_2 \ldots w_{pi} \mathbf{x}_p\\.
\end{equation}

Here, $\mathbf{c}_i$ is the $i^{th}$ principal component. Further, $\mathbf{x}_j \in \mathbb{R}^n$, is the $j^{th}$ column of $\mathbf{X}$ and represents the full set of observations of the $j$th variable, standardized to have zero mean and unit variance. The weights, $w_{ji}$, give the degree to which variable $j$ contributes to component $j$. The weights are selected so that each successive component accounts for the maximum amount of variance possible. As noted in the Online Methods, all of the components, the associated weights (loadings), and the percent variance accounted for by each component can be directly calculated via a singular value decomposition (SVD) of $\mathbf{X}$. Moreover, because each successive component is chosen so as to account for the maximum amount of variance, there exists a single optimal solution.

\begin{figure}[t]
\begin{center}
\centerline{\includegraphics[width=0.5\textwidth]{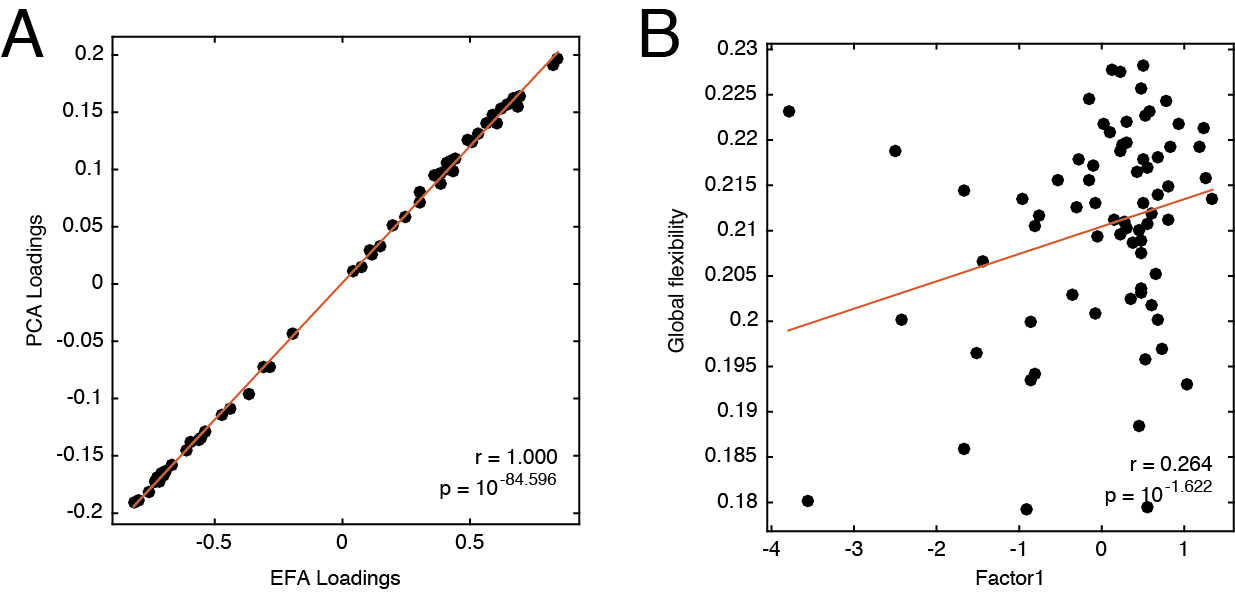}}
\caption{\textbf{Summary of exploratory factor analysis (EFA) with $m = 1$.} \emph{(A)} Correlation of PCA loadings onto $PI$ and loadings of PANAS-X onto the first factor. \emph{(B)} Correlation of factor scores with global flexibility.}\label{figure:figEFA}
\end{center}
\end{figure}

\subsubsection{EFA model}
EFA, on the other hand, is based on the common factor model \cite{thurstone1947multiple} and assumes that each observed variable is a linear combination of unobserved ``common factors'':

\begin{equation}
\mathbf{x}_j = \lambda_{1j} \eta_1 + \lambda_{2j} \eta_2 \ldots \lambda_{mj} \eta_m + \varepsilon_j\\.
\end{equation}

As before, $\mathbf{x}_j$, is the $j^{th}$ observed variable. The variables $\eta_{i}$ represents the unobserved common factors. The weights, $\lambda_{ji}$ give the contribution of factor $i$ to observed variable $j$. Finally, $\varepsilon_j$ is the unique variance associated with variable $j$ (i.e. the variance unaccounted for by the factors). EFA requires the user to specify the number of common factors, $m$. In general, varying $m$ returns different estimates of each factor (i.e. $\eta_1$ with $m=1$ will not be the same as $\eta_1$ if $m>1$).

Unlike PCA, which uses linear algebra to directly calculate components from observed data, EFA estimates both the unique variances \textit{and} the weights (the $\varepsilon$s and $\lambda$s, respectively) in order to model the observed variables. Because EFA amounts to model-fitting, it benefits greatly if there are many times more observations than there are variables; when the ratio of observations to variables is too low, the fitting procedure can lead to unstable parameter estimates \cite{unkel2010simultaneous}.

\subsubsection{EFA model applied to PANAS-X data}
Because the PANAS-X scores have approximately the same number of observations as categories (the actual ratio is $73/57 \approx 1.28$) and because PCA and EFA frequently give comparable results under many practical circumstances, we opted to use PCA in the main text to define an index of positivity. Here, however, we demonstrate that we can derive a similar index using EFA. Specifically, we submit the data matrix of observed PANAS-X scores, $\mathbf{X}$, to an EFA with $m=1$ (i.e. we seek to obtain a single common factor). The result is a set of loadings that are highly consistent with the loadings of PANAS-X scores onto the positivity index (Pearson's correlation, $\hat{r} = 1.00$, $p < 10^{-15}$). Indeed, the top five loadings were ``happy'', ``enthusiastic'', ``confident'', ``cheerful'', and ``delighted.'' The bottom loadings were ``upset'', ``sad'', ``irritable'', ``blue'', and ``downhearted.'' Similarly, when we calculated the correlation of the factor scores with global flexibility, we observed a pattern in line with that described in the main text, where increases in factor scores correspond to increases in global flexibility (Pearson's correlation coefficient, $\hat{r} = 0.264$, $p = 0.024$) (Figure \ref{figure:figEFA}).

\begin{figure}[t]
\begin{center}
\centerline{\includegraphics[width=0.5\textwidth]{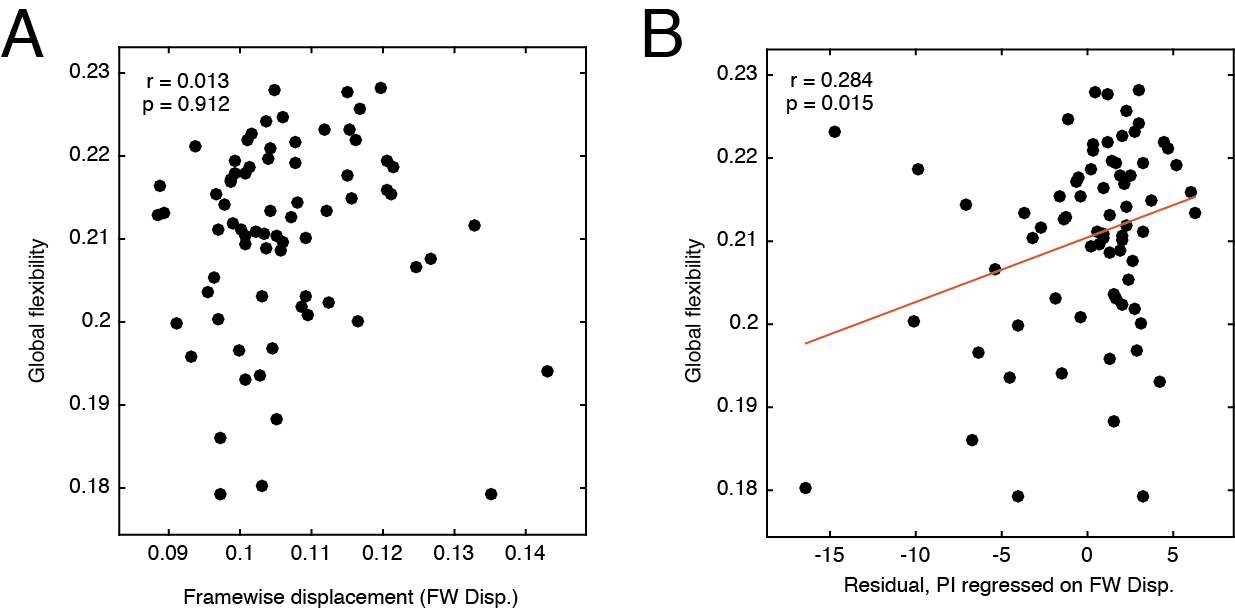}}
\caption{\textbf{Non-significant impact of subject head motion} \emph{(A)} Scatterplot of frame-wise displacement with global flexibility, $F$. \emph{(B)} Scatterplot of global flexibility, $F$, against $PI$ after frame-wise displacement was regressed out.}\label{figure:headMotion}
\end{center}
\end{figure}

\subsection{Nuisance variables}
To this point, we have demonstrated the robustness of the correlation between positivity index, $PI$, and global flexibility, $F$. Another concern is that this relationship is mediated by a third, unmeasured variable. In other words, by virtue of both $PI$ and $F$ being correlated with unknown variable, $x$, we observe a correlation between $PI$ and $F$. To determine whether this was the case, we investigated the relationship of $F$ with head motion and other psychophysiological variables collected as part of the \emph{MyConnectome Project}.

\begin{figure}[t]
\begin{center}
\centerline{\includegraphics[width=0.5\textwidth]{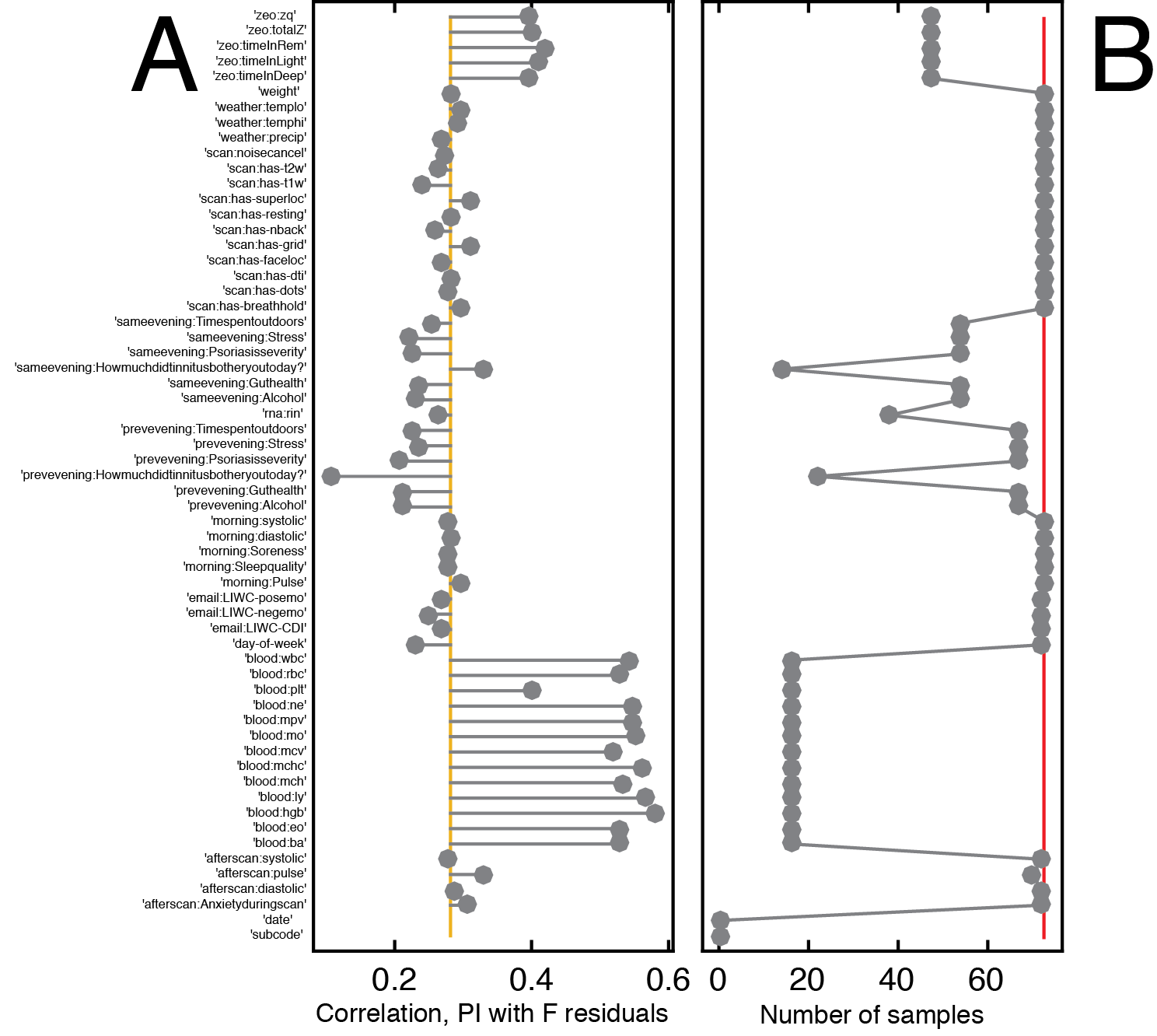}}
\caption{\textbf{Correlation of $PI$ and $F$ after regressing out nuisance variables.} \emph{(A)} Pearson's correlation magnitude of the positivity index, $PI$, with the residuals of $F$ after regressing out each nuisance variable. \emph{(B)} Number of recording sessions over which each nuisance variable was measured.}\label{figure:nuisance}
\end{center}
\end{figure}

\subsubsection{Relationship of global flexibility and head motion}
Recent work has demonstrated that subject head motion within the scanner can introduce systematic biases in functional connectivity patterns \cite{power2012spurious}. It is therefore possible that global flexibility, rather than tracking the reconfiguration of communities over time, is driven by head motion. To test whether this was the case, we asked whether global flexibility was correlated with average frame-wise displacement, which represents an estimate of the amplitude with which a subject's head moves relative to a reference frame. Our analysis revealed no correlation between this motion variable and global flexibility ($\hat{r}(\text{motion},F) = 0.013$, $p=0.912$) (Figure \ref{figure:headMotion}A). In addition, we regressed out frame-wise displacement from the global flexibility estimates and recalculated the correlation of the residuals with the positivity index. This additional step did not influence the magnitude of correlation (which we observed to be $\hat{r}(PI,F) = 0.284$, $p=0.015$) (Figure \ref{figure:headMotion}B).

\subsubsection{Other psychophysiological measurements}
In addition to PANAS-X categories, the \emph{MyConnectome Project} made a number of other psychophysiological measures. For example, on certain recording sessions, blood was drawn and measures such as platelet and red blood cell counts made. Other measures include subjective rates of sleep quality, whether the subject drank alcohol the previous evening, and whether there were precipitation on the day of the scan (See Table\ref{table:nuisance} for a complete list). An important concern is whether these variables account for the correlation between $F$ and $PI$, $\hat{r}(PI,F)$. As with the motion control subsection, we regressed each variable from $F$, and calculated the correlation of the residuals with $PI$. Many of the psychophysiological measurements were not performed in every scan session. For such cases, we performed the regression analysis on the subset of sessions for which those variables were measured. In general, even after controlling for other psychophysiological variables, we still observed a positive correlation between $PI$ and $F$ (Figure \ref{figure:nuisance}A). For most variables, regressing them out from flexibility yielded little change in $\hat{r}(PI,F)$. There are two notable cases, however, that deviate from this trend. First, after controlling for variables based on bloodwork (in \ref{figure:nuisance} they are preceded with the label ``blood:'') we observed a large increase in the magnitude of $\hat{r}(PI,F)$. However, bloodwork was performed on only 16 of the 73 analyzed recording sessions; such a small sample does not permit us to make strong quantitative statements. The second notable case concerned the variable ``how much did tinnitus bother you today?''. Controlling for this variable decreased $\hat{r}(PI,F)$ to $\approx0.1$. Like the bloodwork, this variable was measured during a small fraction of the recording sessions (22 of 73). Again, with such a small sample size we are not in a position to make strong quantitative statements about the relationship of the subject's tinnitus with flexibility. Limiting ourselves to variables that were measured during at least 50\% of the analyzed recording sessions, we found that the median correlation after regressing out each nuisance variable from $F$ was comparable to the correlation magnitude reported in the main text ($\hat{r}(PI,F)_{median} = 0.276$; $\hat{r}(PI,F)_{min} = 0.208$; $\hat{r}(PI,F)_{max} = 0.418$). While not conclusive, these results suggest that the psychophysiological measurements collected in addition to the PANAS-X scores did not, on their own, account for the strength of the correlation of flexibility and the positivity index.

\begin{figure*}[b]
\begin{center}
\centerline{\includegraphics[width=1\textwidth]{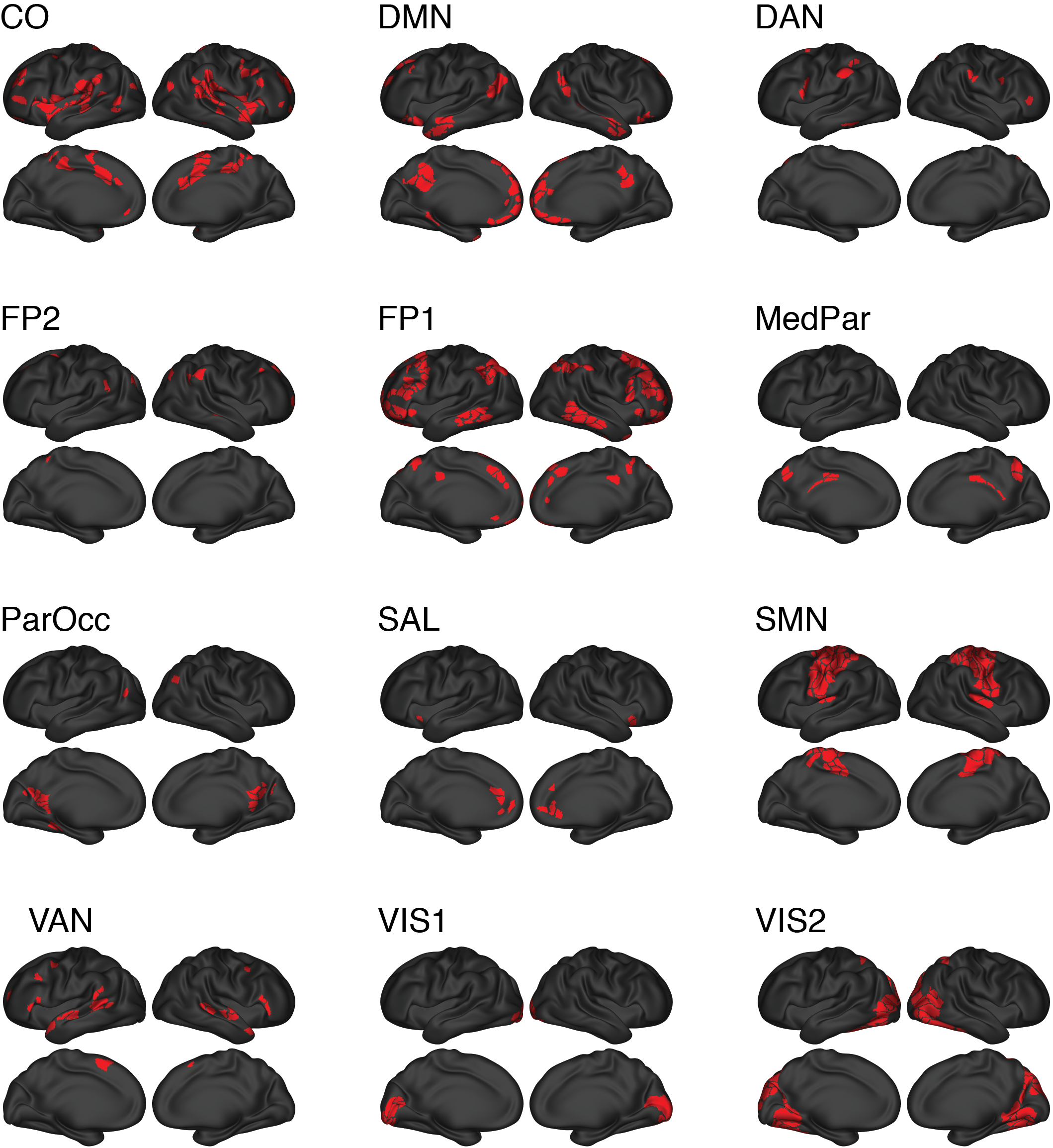}}
\caption{\textbf{Brain systems.} Topographic distribution of the 12 cortical brain systems.}\label{figure:brainSystems}
\end{center}
\end{figure*}

\clearpage

\begin{table*}[h]
\begin{center}
\begin{tabular}{|p{12cm}|}
\hline
active, afraid, alert, alone, amazed, angry, angry-at-self, ashamed, astonished, at-ease, attentive, bashful, blameworthy, blue, bold, calm, cheerful, concentrating, confident, daring, delighted, determined, disgusted, disgusted-with-self, dissatisfied-with-self, distressed, downhearted, drowsy, energetic, enthusiastic, excited, fearless, frightened, guilty, happy, hostile, inspired, interested, irritable, jittery, joyful, lively, loathing, lonely, nervous, proud, relaxed, sad, scared, scornful, shaky, sheepish, shy, sleepy, sluggish, strong, surprised, timid, tired, upset\\ \hline
\end{tabular}
\end{center}
\caption{\textbf{PANAS-X categories.} Complete list of PANAS-X categories.}
\label{table:panasxTerms}
\end{table*}

\begin{table*}[h]
\begin{center}
\begin{tabular}{|l|l|l|}
\hline
\textbf{PANAS-X Class}& \multicolumn{2}{p{12cm}|}{\raggedright \textbf{PANAS-X Category}} \\ \hline \hline
\textbf{negative affect}& \multicolumn{2}{p{12cm}|}{\raggedright {afraid, scared, nervous, jittery, irritable, hostile, guilty, ashamed, upset, distressed}} \\ \hline
\textbf{positive affect}& \multicolumn{2}{p{12cm}|}{\raggedright {active, alert, attentive, determined, enthusiastic, excited, inspired, interested, proud, strong}} \\ \hline
\textbf{fear}& \multicolumn{2}{p{12cm}|}{\raggedright {afraid, scared, frightened, nervous, jittery, shaky}} \\ \hline
\textbf{hostility}& \multicolumn{2}{p{12cm}|}{\raggedright {angry, hostile, irritable, scornful, disgusted, loathing}} \\ \hline
\textbf{guilt}& \multicolumn{2}{p{12cm}|}{\raggedright {guilty, ashamed, blameworthy, angry at self, disgusted with self, dissatisfied with self}} \\ \hline
\textbf{sadness}& \multicolumn{2}{p{12cm}|}{\raggedright {sad, blue, downhearted, alone, lonely}} \\ \hline
\textbf{joviality}& \multicolumn{2}{p{12cm}|}{\raggedright {happy, joyful, delighted, cheerful, excited, enthusiastic, lively, energetic}} \\ \hline
\textbf{self-assurance}& \multicolumn{2}{p{12cm}|}{\raggedright {proud, strong, confident, bold, daring, fearless}} \\ \hline
\textbf{attentiveness}& \multicolumn{2}{p{12cm}|}{\raggedright {alert, attentive, concentrating, determined}} \\ \hline
\textbf{fatigue}& \multicolumn{2}{p{12cm}|}{\raggedright {sleepy, tired, sluggish, drowsy}} \\ \hline
\textbf{serenity}& \multicolumn{2}{p{12cm}|}{\raggedright {calm, relaxed, at ease}} \\ \hline
\textbf{surprise}& \multicolumn{2}{p{12cm}|}{\raggedright {amazed, surprised, astonished}} \\ \hline
\end{tabular}
\end{center}
\caption{\textbf{PANAS-X class assignments.} The first column displays the name of each affect class and the second column displays the PANAS-X scores assigned to that class.}
\label{table:panasxCategories}
\end{table*}

\begin{table*}[h]
{\renewcommand\arraystretch{1.25}
\begin{tabular}{|l|l|l|} \hline
\textbf{Variable class}& \multicolumn{2}{p{12cm}|}{\raggedright \textbf{Variable name}} \\ \hline \hline
\textbf{N/A}:& \multicolumn{2}{l|}{subcode, date} \\ \hline
\textbf{after scan}& \multicolumn{2}{p{12cm}|}{\raggedright anxiety during scan, diastolic, pulse, systolic} \\ \hline
\textbf{blood}& \multicolumn{2}{p{12cm}|}{\raggedright ba, eo, hgb, ly, mch, mchc, mcv, mo, mpv, ne, plt, rbc, wbc} \\ \hline
\textbf{date of week}& \multicolumn{2}{p{12cm}|}{\raggedright date of week} \\ \hline
\textbf{email}& \multicolumn{2}{p{12cm}|}{\raggedright LIWC-CDI, LIWC-negemo, LIWC-posemo} \\ \hline
\textbf{morning}& \multicolumn{2}{p{12cm}|}{\raggedright pulse, sleep quality, soreness, diastolic, systolic} \\ \hline
\textbf{previous evening}& \multicolumn{2}{p{12cm}|}{\raggedright alcohol, gut health, how much did tinnitus bother you today, psoriasis severity, stress, time spent outdoors} \\ \hline
\textbf{rna}& \multicolumn{2}{p{12cm}|}{\raggedright rin} \\ \hline
\textbf{same evening}& \multicolumn{2}{p{12cm}|}{\raggedright alcohol, gut health, how much did tinnitus bother you today, psoriasis severity, stress, time spent outdoors} \\ \hline
\textbf{scan}& \multicolumn{2}{p{12cm}|}{\raggedright has breath hold, has dots, has dti, has faceloc, has grid, has n-back, has resting, has superloc, has T1W, has T2W, noise cancel} \\ \hline
\textbf{weather}& \multicolumn{2}{p{12cm}|}{\raggedright precip, temp hi, temp lo} \\ \hline
\textbf{weight}& \multicolumn{2}{p{12cm}|}{\raggedright weight} \\ \hline
\textbf{zeo}& \multicolumn{2}{p{12cm}|}{\raggedright time in deep, time in light, time in REM, total Z, zq} \\ \hline
\end{tabular}}
\caption{\textbf{Psychophysiological variables} The first column displays the broad class to which each variable was assigned. The second column lists the specific variable names.}
\label{table:nuisance}
\end{table*}

\bibliography{biblio}

\begin{thebibliography}{37}%
\makeatletter
\providecommand \@ifxundefined [1]{%
 \@ifx{#1\undefined}
}%
\providecommand \@ifnum [1]{%
 \ifnum #1\expandafter \@firstoftwo
 \else \expandafter \@secondoftwo
 \fi
}%
\providecommand \@ifx [1]{%
 \ifx #1\expandafter \@firstoftwo
 \else \expandafter \@secondoftwo
 \fi
}%
\providecommand \natexlab [1]{#1}%
\providecommand \enquote  [1]{``#1''}%
\providecommand \bibnamefont  [1]{#1}%
\providecommand \bibfnamefont [1]{#1}%
\providecommand \citenamefont [1]{#1}%
\providecommand \href@noop [0]{\@secondoftwo}%
\providecommand \href [0]{\begingroup \@sanitize@url \@href}%
\providecommand \@href[1]{\@@startlink{#1}\@@href}%
\providecommand \@@href[1]{\endgroup#1\@@endlink}%
\providecommand \@sanitize@url [0]{\catcode `\\12\catcode `\$12\catcode
  `\&12\catcode `\#12\catcode `\^12\catcode `\_12\catcode `\%12\relax}%
\providecommand \@@startlink[1]{}%
\providecommand \@@endlink[0]{}%
\providecommand \url  [0]{\begingroup\@sanitize@url \@url }%
\providecommand \@url [1]{\endgroup\@href {#1}{\urlprefix }}%
\providecommand \urlprefix  [0]{URL }%
\providecommand \Eprint [0]{\href }%
\providecommand \doibase [0]{http://dx.doi.org/}%
\providecommand \selectlanguage [0]{\@gobble}%
\providecommand \bibinfo  [0]{\@secondoftwo}%
\providecommand \bibfield  [0]{\@secondoftwo}%
\providecommand \translation [1]{[#1]}%
\providecommand \BibitemOpen [0]{}%
\providecommand \bibitemStop [0]{}%
\providecommand \bibitemNoStop [0]{.\EOS\space}%
\providecommand \EOS [0]{\spacefactor3000\relax}%
\providecommand \BibitemShut  [1]{\csname bibitem#1\endcsname}%
\let\auto@bib@innerbib\@empty
\bibitem [{\citenamefont {Bassett}\ \emph {et~al.}(2011)\citenamefont
  {Bassett}, \citenamefont {Wymbs}, \citenamefont {Porter}, \citenamefont
  {Mucha}, \citenamefont {Carlson},\ and\ \citenamefont
  {Grafton}}]{bassett2011dynamic}%
  \BibitemOpen
  \bibfield  {author} {\bibinfo {author} {\bibfnamefont {D.~S.}\ \bibnamefont
  {Bassett}}, \bibinfo {author} {\bibfnamefont {N.~F.}\ \bibnamefont {Wymbs}},
  \bibinfo {author} {\bibfnamefont {M.~A.}\ \bibnamefont {Porter}}, \bibinfo
  {author} {\bibfnamefont {P.~J.}\ \bibnamefont {Mucha}}, \bibinfo {author}
  {\bibfnamefont {J.~M.}\ \bibnamefont {Carlson}}, \ and\ \bibinfo {author}
  {\bibfnamefont {S.~T.}\ \bibnamefont {Grafton}},\ }\href@noop {} {\bibfield
  {journal} {\bibinfo  {journal} {Proceedings of the National Academy of
  Sciences}\ }\textbf {\bibinfo {volume} {108}},\ \bibinfo {pages} {7641}
  (\bibinfo {year} {2011})}\BibitemShut {NoStop}%
\bibitem [{\citenamefont {Braun}\ \emph {et~al.}(2015)\citenamefont {Braun},
  \citenamefont {Sch{\"a}fer}, \citenamefont {Walter}, \citenamefont {Erk},
  \citenamefont {Romanczuk-Seiferth}, \citenamefont {Haddad}, \citenamefont
  {Schweiger}, \citenamefont {Grimm}, \citenamefont {Heinz}, \citenamefont
  {Tost} \emph {et~al.}}]{braun2015dynamic}%
  \BibitemOpen
  \bibfield  {author} {\bibinfo {author} {\bibfnamefont {U.}~\bibnamefont
  {Braun}}, \bibinfo {author} {\bibfnamefont {A.}~\bibnamefont {Sch{\"a}fer}},
  \bibinfo {author} {\bibfnamefont {H.}~\bibnamefont {Walter}}, \bibinfo
  {author} {\bibfnamefont {S.}~\bibnamefont {Erk}}, \bibinfo {author}
  {\bibfnamefont {N.}~\bibnamefont {Romanczuk-Seiferth}}, \bibinfo {author}
  {\bibfnamefont {L.}~\bibnamefont {Haddad}}, \bibinfo {author} {\bibfnamefont
  {J.~I.}\ \bibnamefont {Schweiger}}, \bibinfo {author} {\bibfnamefont
  {O.}~\bibnamefont {Grimm}}, \bibinfo {author} {\bibfnamefont
  {A.}~\bibnamefont {Heinz}}, \bibinfo {author} {\bibfnamefont
  {H.}~\bibnamefont {Tost}},  \emph {et~al.},\ }\href@noop {} {\bibfield
  {journal} {\bibinfo  {journal} {Proceedings of the National Academy of
  Sciences}\ }\textbf {\bibinfo {volume} {112}},\ \bibinfo {pages} {11678}
  (\bibinfo {year} {2015})}\BibitemShut {NoStop}%
\bibitem [{\citenamefont {Yerkes}\ and\ \citenamefont
  {Dodson}(1908)}]{robert1908relation}%
  \BibitemOpen
  \bibfield  {author} {\bibinfo {author} {\bibfnamefont {R.~M.}\ \bibnamefont
  {Yerkes}}\ and\ \bibinfo {author} {\bibfnamefont {J.~D.}\ \bibnamefont
  {Dodson}},\ }\href@noop {} {\bibfield  {journal} {\bibinfo  {journal}
  {Journal of comparative neurology and psychology}\ }\textbf {\bibinfo
  {volume} {18}},\ \bibinfo {pages} {459} (\bibinfo {year} {1908})}\BibitemShut
  {NoStop}%
\bibitem [{\citenamefont {Critchley}(2005)}]{critchley2005neural}%
  \BibitemOpen
  \bibfield  {author} {\bibinfo {author} {\bibfnamefont {H.~D.}\ \bibnamefont
  {Critchley}},\ }\href@noop {} {\bibfield  {journal} {\bibinfo  {journal}
  {Journal of Comparative Neurology}\ }\textbf {\bibinfo {volume} {493}},\
  \bibinfo {pages} {154} (\bibinfo {year} {2005})}\BibitemShut {NoStop}%
\bibitem [{\citenamefont {Eldar}\ \emph {et~al.}(2016)\citenamefont {Eldar},
  \citenamefont {Rutledge}, \citenamefont {Dolan},\ and\ \citenamefont
  {Niv}}]{eldar2016mood}%
  \BibitemOpen
  \bibfield  {author} {\bibinfo {author} {\bibfnamefont {E.}~\bibnamefont
  {Eldar}}, \bibinfo {author} {\bibfnamefont {R.~B.}\ \bibnamefont {Rutledge}},
  \bibinfo {author} {\bibfnamefont {R.~J.}\ \bibnamefont {Dolan}}, \ and\
  \bibinfo {author} {\bibfnamefont {Y.}~\bibnamefont {Niv}},\ }\href@noop {}
  {\bibfield  {journal} {\bibinfo  {journal} {Trends in Cognitive Sciences}\
  }\textbf {\bibinfo {volume} {20}},\ \bibinfo {pages} {15} (\bibinfo {year}
  {2016})}\BibitemShut {NoStop}%
\bibitem [{\citenamefont {Nassar}\ \emph {et~al.}(2012)\citenamefont {Nassar},
  \citenamefont {Rumsey}, \citenamefont {Wilson}, \citenamefont {Parikh},
  \citenamefont {Heasly},\ and\ \citenamefont {Gold}}]{nassar2012rational}%
  \BibitemOpen
  \bibfield  {author} {\bibinfo {author} {\bibfnamefont {M.~R.}\ \bibnamefont
  {Nassar}}, \bibinfo {author} {\bibfnamefont {K.~M.}\ \bibnamefont {Rumsey}},
  \bibinfo {author} {\bibfnamefont {R.~C.}\ \bibnamefont {Wilson}}, \bibinfo
  {author} {\bibfnamefont {K.}~\bibnamefont {Parikh}}, \bibinfo {author}
  {\bibfnamefont {B.}~\bibnamefont {Heasly}}, \ and\ \bibinfo {author}
  {\bibfnamefont {J.~I.}\ \bibnamefont {Gold}},\ }\href@noop {} {\bibfield
  {journal} {\bibinfo  {journal} {Nature Neuroscience}\ }\textbf {\bibinfo
  {volume} {15}},\ \bibinfo {pages} {1040} (\bibinfo {year}
  {2012})}\BibitemShut {NoStop}%
\bibitem [{\citenamefont {Laumann}\ \emph {et~al.}(2015)\citenamefont
  {Laumann}, \citenamefont {Gordon}, \citenamefont {Adeyemo}, \citenamefont
  {Snyder}, \citenamefont {Joo}, \citenamefont {Chen}, \citenamefont {Gilmore},
  \citenamefont {McDermott}, \citenamefont {Nelson}, \citenamefont {Dosenbach}
  \emph {et~al.}}]{laumann2015functional}%
  \BibitemOpen
  \bibfield  {author} {\bibinfo {author} {\bibfnamefont {T.~O.}\ \bibnamefont
  {Laumann}}, \bibinfo {author} {\bibfnamefont {E.~M.}\ \bibnamefont {Gordon}},
  \bibinfo {author} {\bibfnamefont {B.}~\bibnamefont {Adeyemo}}, \bibinfo
  {author} {\bibfnamefont {A.~Z.}\ \bibnamefont {Snyder}}, \bibinfo {author}
  {\bibfnamefont {S.~J.}\ \bibnamefont {Joo}}, \bibinfo {author} {\bibfnamefont
  {M.-Y.}\ \bibnamefont {Chen}}, \bibinfo {author} {\bibfnamefont {A.~W.}\
  \bibnamefont {Gilmore}}, \bibinfo {author} {\bibfnamefont {K.~B.}\
  \bibnamefont {McDermott}}, \bibinfo {author} {\bibfnamefont {S.~M.}\
  \bibnamefont {Nelson}}, \bibinfo {author} {\bibfnamefont {N.~U.}\
  \bibnamefont {Dosenbach}},  \emph {et~al.},\ }\href@noop {} {\bibfield
  {journal} {\bibinfo  {journal} {Neuron}\ }\textbf {\bibinfo {volume} {87}},\
  \bibinfo {pages} {657} (\bibinfo {year} {2015})}\BibitemShut {NoStop}%
\bibitem [{\citenamefont {Poldrack}\ \emph {et~al.}(2015)\citenamefont
  {Poldrack}, \citenamefont {Laumann}, \citenamefont {Koyejo}, \citenamefont
  {Gregory}, \citenamefont {Hover}, \citenamefont {Chen}, \citenamefont
  {Gorgolewski}, \citenamefont {Luci}, \citenamefont {Joo}, \citenamefont
  {Boyd} \emph {et~al.}}]{poldrack2015long}%
  \BibitemOpen
  \bibfield  {author} {\bibinfo {author} {\bibfnamefont {R.~A.}\ \bibnamefont
  {Poldrack}}, \bibinfo {author} {\bibfnamefont {T.~O.}\ \bibnamefont
  {Laumann}}, \bibinfo {author} {\bibfnamefont {O.}~\bibnamefont {Koyejo}},
  \bibinfo {author} {\bibfnamefont {B.}~\bibnamefont {Gregory}}, \bibinfo
  {author} {\bibfnamefont {A.}~\bibnamefont {Hover}}, \bibinfo {author}
  {\bibfnamefont {M.-Y.}\ \bibnamefont {Chen}}, \bibinfo {author}
  {\bibfnamefont {K.~J.}\ \bibnamefont {Gorgolewski}}, \bibinfo {author}
  {\bibfnamefont {J.}~\bibnamefont {Luci}}, \bibinfo {author} {\bibfnamefont
  {S.~J.}\ \bibnamefont {Joo}}, \bibinfo {author} {\bibfnamefont {R.~L.}\
  \bibnamefont {Boyd}},  \emph {et~al.},\ }\href@noop {} {\bibfield  {journal}
  {\bibinfo  {journal} {Nature communications}\ }\textbf {\bibinfo {volume}
  {6}} (\bibinfo {year} {2015})}\BibitemShut {NoStop}%
\bibitem [{\citenamefont {Watson}\ \emph {et~al.}(1988)\citenamefont {Watson},
  \citenamefont {Clark},\ and\ \citenamefont
  {Tellegen}}]{watson1988development}%
  \BibitemOpen
  \bibfield  {author} {\bibinfo {author} {\bibfnamefont {D.}~\bibnamefont
  {Watson}}, \bibinfo {author} {\bibfnamefont {L.~A.}\ \bibnamefont {Clark}}, \
  and\ \bibinfo {author} {\bibfnamefont {A.}~\bibnamefont {Tellegen}},\
  }\href@noop {} {\bibfield  {journal} {\bibinfo  {journal} {Journal of
  Personality and Social Psychology}\ }\textbf {\bibinfo {volume} {54}},\
  \bibinfo {pages} {1063} (\bibinfo {year} {1988})}\BibitemShut {NoStop}%
\bibitem [{\citenamefont {Bassett}\ \emph {et~al.}(2013)\citenamefont
  {Bassett}, \citenamefont {Porter}, \citenamefont {Wymbs}, \citenamefont
  {Grafton}, \citenamefont {Carlson},\ and\ \citenamefont
  {Mucha}}]{bassett2013robust}%
  \BibitemOpen
  \bibfield  {author} {\bibinfo {author} {\bibfnamefont {D.~S.}\ \bibnamefont
  {Bassett}}, \bibinfo {author} {\bibfnamefont {M.~A.}\ \bibnamefont {Porter}},
  \bibinfo {author} {\bibfnamefont {N.~F.}\ \bibnamefont {Wymbs}}, \bibinfo
  {author} {\bibfnamefont {S.~T.}\ \bibnamefont {Grafton}}, \bibinfo {author}
  {\bibfnamefont {J.~M.}\ \bibnamefont {Carlson}}, \ and\ \bibinfo {author}
  {\bibfnamefont {P.~J.}\ \bibnamefont {Mucha}},\ }\href@noop {} {\bibfield
  {journal} {\bibinfo  {journal} {Chaos: An Interdisciplinary Journal of
  Nonlinear Science}\ }\textbf {\bibinfo {volume} {23}},\ \bibinfo {pages}
  {013142} (\bibinfo {year} {2013})}\BibitemShut {NoStop}%
\bibitem [{\citenamefont {Power}\ \emph {et~al.}(2011)\citenamefont {Power},
  \citenamefont {Cohen}, \citenamefont {Nelson}, \citenamefont {Wig},
  \citenamefont {Barnes}, \citenamefont {Church}, \citenamefont {Vogel},
  \citenamefont {Laumann}, \citenamefont {Miezin}, \citenamefont {Schlaggar}
  \emph {et~al.}}]{power2011functional}%
  \BibitemOpen
  \bibfield  {author} {\bibinfo {author} {\bibfnamefont {J.~D.}\ \bibnamefont
  {Power}}, \bibinfo {author} {\bibfnamefont {A.~L.}\ \bibnamefont {Cohen}},
  \bibinfo {author} {\bibfnamefont {S.~M.}\ \bibnamefont {Nelson}}, \bibinfo
  {author} {\bibfnamefont {G.~S.}\ \bibnamefont {Wig}}, \bibinfo {author}
  {\bibfnamefont {K.~A.}\ \bibnamefont {Barnes}}, \bibinfo {author}
  {\bibfnamefont {J.~A.}\ \bibnamefont {Church}}, \bibinfo {author}
  {\bibfnamefont {A.~C.}\ \bibnamefont {Vogel}}, \bibinfo {author}
  {\bibfnamefont {T.~O.}\ \bibnamefont {Laumann}}, \bibinfo {author}
  {\bibfnamefont {F.~M.}\ \bibnamefont {Miezin}}, \bibinfo {author}
  {\bibfnamefont {B.~L.}\ \bibnamefont {Schlaggar}},  \emph {et~al.},\
  }\href@noop {} {\bibfield  {journal} {\bibinfo  {journal} {Neuron}\ }\textbf
  {\bibinfo {volume} {72}},\ \bibinfo {pages} {665} (\bibinfo {year}
  {2011})}\BibitemShut {NoStop}%
\bibitem [{\citenamefont {Mucha}\ \emph {et~al.}(2010)\citenamefont {Mucha},
  \citenamefont {Richardson}, \citenamefont {Macon}, \citenamefont {Porter},\
  and\ \citenamefont {Onnela}}]{mucha2010community}%
  \BibitemOpen
  \bibfield  {author} {\bibinfo {author} {\bibfnamefont {P.~J.}\ \bibnamefont
  {Mucha}}, \bibinfo {author} {\bibfnamefont {T.}~\bibnamefont {Richardson}},
  \bibinfo {author} {\bibfnamefont {K.}~\bibnamefont {Macon}}, \bibinfo
  {author} {\bibfnamefont {M.~A.}\ \bibnamefont {Porter}}, \ and\ \bibinfo
  {author} {\bibfnamefont {J.-P.}\ \bibnamefont {Onnela}},\ }\href@noop {}
  {\bibfield  {journal} {\bibinfo  {journal} {Science}\ }\textbf {\bibinfo
  {volume} {328}},\ \bibinfo {pages} {876} (\bibinfo {year}
  {2010})}\BibitemShut {NoStop}%
\bibitem [{\citenamefont {Adolphs}(2002)}]{adolphs2002neural}%
  \BibitemOpen
  \bibfield  {author} {\bibinfo {author} {\bibfnamefont {R.}~\bibnamefont
  {Adolphs}},\ }\href@noop {} {\bibfield  {journal} {\bibinfo  {journal}
  {Current opinion in neurobiology}\ }\textbf {\bibinfo {volume} {12}},\
  \bibinfo {pages} {169} (\bibinfo {year} {2002})}\BibitemShut {NoStop}%
\bibitem [{\citenamefont {Chrobak}\ \emph {et~al.}(2015)\citenamefont
  {Chrobak}, \citenamefont {Siuda-Krzywicka}, \citenamefont {Siwek},
  \citenamefont {Arciszewska}, \citenamefont {Siwek}, \citenamefont
  {Starowicz-Filip},\ and\ \citenamefont {Dudek}}]{chrobak2015implicit}%
  \BibitemOpen
  \bibfield  {author} {\bibinfo {author} {\bibfnamefont {A.~A.}\ \bibnamefont
  {Chrobak}}, \bibinfo {author} {\bibfnamefont {K.}~\bibnamefont
  {Siuda-Krzywicka}}, \bibinfo {author} {\bibfnamefont {G.~P.}\ \bibnamefont
  {Siwek}}, \bibinfo {author} {\bibfnamefont {A.}~\bibnamefont {Arciszewska}},
  \bibinfo {author} {\bibfnamefont {M.}~\bibnamefont {Siwek}}, \bibinfo
  {author} {\bibfnamefont {A.}~\bibnamefont {Starowicz-Filip}}, \ and\ \bibinfo
  {author} {\bibfnamefont {D.}~\bibnamefont {Dudek}},\ }\href@noop {}
  {\bibfield  {journal} {\bibinfo  {journal} {Journal of Affective Disorders}\
  }\textbf {\bibinfo {volume} {174}},\ \bibinfo {pages} {250} (\bibinfo {year}
  {2015})}\BibitemShut {NoStop}%
\bibitem [{\citenamefont {Papmeyer}\ \emph {et~al.}(2015)\citenamefont
  {Papmeyer}, \citenamefont {Sussmann}, \citenamefont {Hall}, \citenamefont
  {McKirdy}, \citenamefont {Peel}, \citenamefont {Macdonald}, \citenamefont
  {Lawrie}, \citenamefont {Whalley},\ and\ \citenamefont
  {McIntosh}}]{papmeyer2015neurocognition}%
  \BibitemOpen
  \bibfield  {author} {\bibinfo {author} {\bibfnamefont {M.}~\bibnamefont
  {Papmeyer}}, \bibinfo {author} {\bibfnamefont {J.}~\bibnamefont {Sussmann}},
  \bibinfo {author} {\bibfnamefont {J.}~\bibnamefont {Hall}}, \bibinfo {author}
  {\bibfnamefont {J.}~\bibnamefont {McKirdy}}, \bibinfo {author} {\bibfnamefont
  {A.}~\bibnamefont {Peel}}, \bibinfo {author} {\bibfnamefont {A.}~\bibnamefont
  {Macdonald}}, \bibinfo {author} {\bibfnamefont {S.}~\bibnamefont {Lawrie}},
  \bibinfo {author} {\bibfnamefont {H.}~\bibnamefont {Whalley}}, \ and\
  \bibinfo {author} {\bibfnamefont {A.}~\bibnamefont {McIntosh}},\ }\href@noop
  {} {\bibfield  {journal} {\bibinfo  {journal} {Psychological Medicine}\
  }\textbf {\bibinfo {volume} {45}},\ \bibinfo {pages} {3317} (\bibinfo {year}
  {2015})}\BibitemShut {NoStop}%
\bibitem [{\citenamefont {Braun}\ \emph {et~al.}(2016)\citenamefont {Braun},
  \citenamefont {Sch{\"a}fer}, \citenamefont {Bassett}, \citenamefont {Rausch},
  \citenamefont {Schweiger}, \citenamefont {Bilek}, \citenamefont {Erk},
  \citenamefont {Romanczuk-Seiferth}, \citenamefont {Grimm}, \citenamefont
  {Haddad}, \citenamefont {Otto}, \citenamefont {Mohnke}, \citenamefont
  {Heinz}, \citenamefont {Zink}, \citenamefont {Walter}, \citenamefont
  {Meyer-Lindeberg},\ and\ \citenamefont {Tost}}]{braun2016dynamic}%
  \BibitemOpen
  \bibfield  {author} {\bibinfo {author} {\bibfnamefont {U.}~\bibnamefont
  {Braun}}, \bibinfo {author} {\bibfnamefont {A.}~\bibnamefont {Sch{\"a}fer}},
  \bibinfo {author} {\bibfnamefont {D.~S.}\ \bibnamefont {Bassett}}, \bibinfo
  {author} {\bibfnamefont {F.}~\bibnamefont {Rausch}}, \bibinfo {author}
  {\bibfnamefont {J.}~\bibnamefont {Schweiger}}, \bibinfo {author}
  {\bibfnamefont {E.}~\bibnamefont {Bilek}}, \bibinfo {author} {\bibfnamefont
  {S.}~\bibnamefont {Erk}}, \bibinfo {author} {\bibfnamefont {N.}~\bibnamefont
  {Romanczuk-Seiferth}}, \bibinfo {author} {\bibfnamefont {O.}~\bibnamefont
  {Grimm}}, \bibinfo {author} {\bibfnamefont {L.}~\bibnamefont {Haddad}},
  \bibinfo {author} {\bibfnamefont {K.}~\bibnamefont {Otto}}, \bibinfo {author}
  {\bibfnamefont {S.}~\bibnamefont {Mohnke}}, \bibinfo {author} {\bibfnamefont
  {A.}~\bibnamefont {Heinz}}, \bibinfo {author} {\bibfnamefont
  {M.}~\bibnamefont {Zink}}, \bibinfo {author} {\bibfnamefont {H.}~\bibnamefont
  {Walter}}, \bibinfo {author} {\bibfnamefont {A.}~\bibnamefont
  {Meyer-Lindeberg}}, \ and\ \bibinfo {author} {\bibfnamefont {H.}~\bibnamefont
  {Tost}},\ }\href@noop {} {\bibfield  {journal} {\bibinfo  {journal}
  {submitted}\ } (\bibinfo {year} {2016})}\BibitemShut {NoStop}%
\bibitem [{\citenamefont {Hegerl}\ and\ \citenamefont
  {Hensch}(2014)}]{hegerl2014vigilance}%
  \BibitemOpen
  \bibfield  {author} {\bibinfo {author} {\bibfnamefont {U.}~\bibnamefont
  {Hegerl}}\ and\ \bibinfo {author} {\bibfnamefont {T.}~\bibnamefont
  {Hensch}},\ }\href@noop {} {\bibfield  {journal} {\bibinfo  {journal}
  {Neuroscience \& Biobehavioral Reviews}\ }\textbf {\bibinfo {volume} {44}},\
  \bibinfo {pages} {45} (\bibinfo {year} {2014})}\BibitemShut {NoStop}%
\bibitem [{\citenamefont {Devonshire}\ and\ \citenamefont
  {Dommett}(2010)}]{devonshire2010neuroscience}%
  \BibitemOpen
  \bibfield  {author} {\bibinfo {author} {\bibfnamefont {I.~M.}\ \bibnamefont
  {Devonshire}}\ and\ \bibinfo {author} {\bibfnamefont {E.~J.}\ \bibnamefont
  {Dommett}},\ }\href@noop {} {\bibfield  {journal} {\bibinfo  {journal}
  {Neuroscientist}\ }\textbf {\bibinfo {volume} {16}},\ \bibinfo {pages} {349}
  (\bibinfo {year} {2010})}\BibitemShut {NoStop}%
\bibitem [{\citenamefont {Cascio}\ \emph {et~al.}(2015)\citenamefont {Cascio},
  \citenamefont {O’Donnell}, \citenamefont {Tinney}, \citenamefont
  {Lieberman}, \citenamefont {Taylor}, \citenamefont {Strecher},\ and\
  \citenamefont {Falk}}]{cascio2015self}%
  \BibitemOpen
  \bibfield  {author} {\bibinfo {author} {\bibfnamefont {C.~N.}\ \bibnamefont
  {Cascio}}, \bibinfo {author} {\bibfnamefont {M.~B.}\ \bibnamefont
  {O’Donnell}}, \bibinfo {author} {\bibfnamefont {F.~J.}\ \bibnamefont
  {Tinney}}, \bibinfo {author} {\bibfnamefont {M.~D.}\ \bibnamefont
  {Lieberman}}, \bibinfo {author} {\bibfnamefont {S.~E.}\ \bibnamefont
  {Taylor}}, \bibinfo {author} {\bibfnamefont {V.~J.}\ \bibnamefont
  {Strecher}}, \ and\ \bibinfo {author} {\bibfnamefont {E.~B.}\ \bibnamefont
  {Falk}},\ }\href@noop {} {\bibfield  {journal} {\bibinfo  {journal} {Social
  Cognitive and Affective Neuroscience}\ ,\ \bibinfo {pages} {nsv136}}
  (\bibinfo {year} {2015})}\BibitemShut {NoStop}%
\bibitem [{\citenamefont {Falk}\ \emph {et~al.}(2015)\citenamefont {Falk},
  \citenamefont {O’Donnell}, \citenamefont {Cascio}, \citenamefont {Tinney},
  \citenamefont {Kang}, \citenamefont {Lieberman}, \citenamefont {Taylor},
  \citenamefont {An}, \citenamefont {Resnicow},\ and\ \citenamefont
  {Strecher}}]{falk2015self}%
  \BibitemOpen
  \bibfield  {author} {\bibinfo {author} {\bibfnamefont {E.~B.}\ \bibnamefont
  {Falk}}, \bibinfo {author} {\bibfnamefont {M.~B.}\ \bibnamefont
  {O’Donnell}}, \bibinfo {author} {\bibfnamefont {C.~N.}\ \bibnamefont
  {Cascio}}, \bibinfo {author} {\bibfnamefont {F.}~\bibnamefont {Tinney}},
  \bibinfo {author} {\bibfnamefont {Y.}~\bibnamefont {Kang}}, \bibinfo {author}
  {\bibfnamefont {M.~D.}\ \bibnamefont {Lieberman}}, \bibinfo {author}
  {\bibfnamefont {S.~E.}\ \bibnamefont {Taylor}}, \bibinfo {author}
  {\bibfnamefont {L.}~\bibnamefont {An}}, \bibinfo {author} {\bibfnamefont
  {K.}~\bibnamefont {Resnicow}}, \ and\ \bibinfo {author} {\bibfnamefont
  {V.~J.}\ \bibnamefont {Strecher}},\ }\href@noop {} {\bibfield  {journal}
  {\bibinfo  {journal} {Proceedings of the National Academy of Sciences}\
  }\textbf {\bibinfo {volume} {112}},\ \bibinfo {pages} {1977} (\bibinfo {year}
  {2015})}\BibitemShut {NoStop}%
\bibitem [{\citenamefont {Eckart}\ and\ \citenamefont
  {Young}(1936)}]{eckart1936approximation}%
  \BibitemOpen
  \bibfield  {author} {\bibinfo {author} {\bibfnamefont {C.}~\bibnamefont
  {Eckart}}\ and\ \bibinfo {author} {\bibfnamefont {G.}~\bibnamefont {Young}},\
  }\href@noop {} {\bibfield  {journal} {\bibinfo  {journal} {Psychometrika}\
  }\textbf {\bibinfo {volume} {1}},\ \bibinfo {pages} {211} (\bibinfo {year}
  {1936})}\BibitemShut {NoStop}%
\bibitem [{\citenamefont {Sporns}\ and\ \citenamefont
  {Betzel}(2016)}]{sporns2016modular}%
  \BibitemOpen
  \bibfield  {author} {\bibinfo {author} {\bibfnamefont {O.}~\bibnamefont
  {Sporns}}\ and\ \bibinfo {author} {\bibfnamefont {R.~F.}\ \bibnamefont
  {Betzel}},\ }\href@noop {} {\bibfield  {journal} {\bibinfo  {journal} {Annual
  review of psychology}\ }\textbf {\bibinfo {volume} {67}} (\bibinfo {year}
  {2016})}\BibitemShut {NoStop}%
\bibitem [{\citenamefont {Jutla}\ \emph {et~al.}(2011)\citenamefont {Jutla},
  \citenamefont {Jeub},\ and\ \citenamefont {Mucha}}]{jutla2011generalized}%
  \BibitemOpen
  \bibfield  {author} {\bibinfo {author} {\bibfnamefont {I.~S.}\ \bibnamefont
  {Jutla}}, \bibinfo {author} {\bibfnamefont {L.~G.}\ \bibnamefont {Jeub}}, \
  and\ \bibinfo {author} {\bibfnamefont {P.~J.}\ \bibnamefont {Mucha}},\
  }\href@noop {} {\bibfield  {journal} {\bibinfo  {journal} {URL
  http://netwiki. amath. unc. edu/GenLouvain}\ } (\bibinfo {year}
  {2011})}\BibitemShut {NoStop}%
\bibitem [{\citenamefont {Good}\ \emph {et~al.}(2010)\citenamefont {Good},
  \citenamefont {de~Montjoye},\ and\ \citenamefont
  {Clauset}}]{good2010performance}%
  \BibitemOpen
  \bibfield  {author} {\bibinfo {author} {\bibfnamefont {B.~H.}\ \bibnamefont
  {Good}}, \bibinfo {author} {\bibfnamefont {Y.-A.}\ \bibnamefont
  {de~Montjoye}}, \ and\ \bibinfo {author} {\bibfnamefont {A.}~\bibnamefont
  {Clauset}},\ }\href@noop {} {\bibfield  {journal} {\bibinfo  {journal}
  {Physical Review E}\ }\textbf {\bibinfo {volume} {81}},\ \bibinfo {pages}
  {046106} (\bibinfo {year} {2010})}\BibitemShut {NoStop}%
\bibitem [{\citenamefont {Blondel}\ \emph {et~al.}(2008)\citenamefont
  {Blondel}, \citenamefont {Guillaume}, \citenamefont {Lambiotte},\ and\
  \citenamefont {Lefebvre}}]{blondel2008fast}%
  \BibitemOpen
  \bibfield  {author} {\bibinfo {author} {\bibfnamefont {V.~D.}\ \bibnamefont
  {Blondel}}, \bibinfo {author} {\bibfnamefont {J.-L.}\ \bibnamefont
  {Guillaume}}, \bibinfo {author} {\bibfnamefont {R.}~\bibnamefont
  {Lambiotte}}, \ and\ \bibinfo {author} {\bibfnamefont {E.}~\bibnamefont
  {Lefebvre}},\ }\href@noop {} {\bibfield  {journal} {\bibinfo  {journal}
  {Journal of Statistical Mechanics: Theory and Experiment}\ }\textbf {\bibinfo
  {volume} {2008}},\ \bibinfo {pages} {P10008} (\bibinfo {year}
  {2008})}\BibitemShut {NoStop}%
\bibitem [{\citenamefont {Benjamini}\ and\ \citenamefont
  {Hochberg}(1995)}]{benjamini1995controlling}%
  \BibitemOpen
  \bibfield  {author} {\bibinfo {author} {\bibfnamefont {Y.}~\bibnamefont
  {Benjamini}}\ and\ \bibinfo {author} {\bibfnamefont {Y.}~\bibnamefont
  {Hochberg}},\ }\href@noop {} {\bibfield  {journal} {\bibinfo  {journal}
  {Journal of the Royal Statistical Society. Series B (Methodological)}\ ,\
  \bibinfo {pages} {289}} (\bibinfo {year} {1995})}\BibitemShut {NoStop}%
\bibitem [{\citenamefont {Newman}(2012)}]{newman2012communities}%
  \BibitemOpen
  \bibfield  {author} {\bibinfo {author} {\bibfnamefont {M.~E.}\ \bibnamefont
  {Newman}},\ }\href@noop {} {\bibfield  {journal} {\bibinfo  {journal} {Nature
  Physics}\ }\textbf {\bibinfo {volume} {8}},\ \bibinfo {pages} {25} (\bibinfo
  {year} {2012})}\BibitemShut {NoStop}%
\bibitem [{\citenamefont {Fortunato}(2010)}]{fortunato2010community}%
  \BibitemOpen
  \bibfield  {author} {\bibinfo {author} {\bibfnamefont {S.}~\bibnamefont
  {Fortunato}},\ }\href@noop {} {\bibfield  {journal} {\bibinfo  {journal}
  {Physics Reports}\ }\textbf {\bibinfo {volume} {486}},\ \bibinfo {pages} {75}
  (\bibinfo {year} {2010})}\BibitemShut {NoStop}%
\bibitem [{\citenamefont {Newman}\ and\ \citenamefont
  {Girvan}(2004)}]{newman2004finding}%
  \BibitemOpen
  \bibfield  {author} {\bibinfo {author} {\bibfnamefont {M.~E.}\ \bibnamefont
  {Newman}}\ and\ \bibinfo {author} {\bibfnamefont {M.}~\bibnamefont
  {Girvan}},\ }\href@noop {} {\bibfield  {journal} {\bibinfo  {journal}
  {Physical review E}\ }\textbf {\bibinfo {volume} {69}},\ \bibinfo {pages}
  {026113} (\bibinfo {year} {2004})}\BibitemShut {NoStop}%
\bibitem [{\citenamefont {Kivel{\"a}}\ \emph {et~al.}(2014)\citenamefont
  {Kivel{\"a}}, \citenamefont {Arenas}, \citenamefont {Barthelemy},
  \citenamefont {Gleeson}, \citenamefont {Moreno},\ and\ \citenamefont
  {Porter}}]{kivela2014multilayer}%
  \BibitemOpen
  \bibfield  {author} {\bibinfo {author} {\bibfnamefont {M.}~\bibnamefont
  {Kivel{\"a}}}, \bibinfo {author} {\bibfnamefont {A.}~\bibnamefont {Arenas}},
  \bibinfo {author} {\bibfnamefont {M.}~\bibnamefont {Barthelemy}}, \bibinfo
  {author} {\bibfnamefont {J.~P.}\ \bibnamefont {Gleeson}}, \bibinfo {author}
  {\bibfnamefont {Y.}~\bibnamefont {Moreno}}, \ and\ \bibinfo {author}
  {\bibfnamefont {M.~A.}\ \bibnamefont {Porter}},\ }\href@noop {} {\bibfield
  {journal} {\bibinfo  {journal} {Journal of Complex Networks}\ }\textbf
  {\bibinfo {volume} {2}},\ \bibinfo {pages} {203} (\bibinfo {year}
  {2014})}\BibitemShut {NoStop}%
\bibitem [{\citenamefont {Watson}\ and\ \citenamefont
  {Clark}(1999)}]{watson1999panas}%
  \BibitemOpen
  \bibfield  {author} {\bibinfo {author} {\bibfnamefont {D.}~\bibnamefont
  {Watson}}\ and\ \bibinfo {author} {\bibfnamefont {L.~A.}\ \bibnamefont
  {Clark}},\ }\href@noop {} {\  (\bibinfo {year} {1999})}\BibitemShut {NoStop}%
\bibitem [{\citenamefont {Comrey}(1988)}]{comrey1988factor}%
  \BibitemOpen
  \bibfield  {author} {\bibinfo {author} {\bibfnamefont {A.~L.}\ \bibnamefont
  {Comrey}},\ }\href@noop {} {\bibfield  {journal} {\bibinfo  {journal}
  {Journal of consulting and clinical psychology}\ }\textbf {\bibinfo {volume}
  {56}},\ \bibinfo {pages} {754} (\bibinfo {year} {1988})}\BibitemShut
  {NoStop}%
\bibitem [{\citenamefont {Galbraith}\ \emph {et~al.}(2002)\citenamefont
  {Galbraith}, \citenamefont {Moustaki}, \citenamefont {Bartholomew},\ and\
  \citenamefont {Steele}}]{galbraith2002analysis}%
  \BibitemOpen
  \bibfield  {author} {\bibinfo {author} {\bibfnamefont {J.}~\bibnamefont
  {Galbraith}}, \bibinfo {author} {\bibfnamefont {I.}~\bibnamefont {Moustaki}},
  \bibinfo {author} {\bibfnamefont {D.~J.}\ \bibnamefont {Bartholomew}}, \ and\
  \bibinfo {author} {\bibfnamefont {F.}~\bibnamefont {Steele}},\ }\href@noop {}
  {\emph {\bibinfo {title} {The analysis and interpretation of multivariate
  data for social scientists}}}\ (\bibinfo  {publisher} {CRC Press},\ \bibinfo
  {year} {2002})\BibitemShut {NoStop}%
\bibitem [{\citenamefont {Fabrigar}\ \emph {et~al.}(1999)\citenamefont
  {Fabrigar}, \citenamefont {Wegener}, \citenamefont {MacCallum},\ and\
  \citenamefont {Strahan}}]{fabrigar1999evaluating}%
  \BibitemOpen
  \bibfield  {author} {\bibinfo {author} {\bibfnamefont {L.~R.}\ \bibnamefont
  {Fabrigar}}, \bibinfo {author} {\bibfnamefont {D.~T.}\ \bibnamefont
  {Wegener}}, \bibinfo {author} {\bibfnamefont {R.~C.}\ \bibnamefont
  {MacCallum}}, \ and\ \bibinfo {author} {\bibfnamefont {E.~J.}\ \bibnamefont
  {Strahan}},\ }\href@noop {} {\bibfield  {journal} {\bibinfo  {journal}
  {Psychological methods}\ }\textbf {\bibinfo {volume} {4}},\ \bibinfo {pages}
  {272} (\bibinfo {year} {1999})}\BibitemShut {NoStop}%
\bibitem [{\citenamefont {Thurstone}(1947)}]{thurstone1947multiple}%
  \BibitemOpen
  \bibfield  {author} {\bibinfo {author} {\bibfnamefont {L.~L.}\ \bibnamefont
  {Thurstone}},\ }\href@noop {} {\  (\bibinfo {year} {1947})}\BibitemShut
  {NoStop}%
\bibitem [{\citenamefont {Unkel}\ and\ \citenamefont
  {Trendafilov}(2010)}]{unkel2010simultaneous}%
  \BibitemOpen
  \bibfield  {author} {\bibinfo {author} {\bibfnamefont {S.}~\bibnamefont
  {Unkel}}\ and\ \bibinfo {author} {\bibfnamefont {N.~T.}\ \bibnamefont
  {Trendafilov}},\ }\href@noop {} {\bibfield  {journal} {\bibinfo  {journal}
  {International Statistical Review}\ }\textbf {\bibinfo {volume} {78}},\
  \bibinfo {pages} {363} (\bibinfo {year} {2010})}\BibitemShut {NoStop}%
\bibitem [{\citenamefont {Power}\ \emph {et~al.}(2012)\citenamefont {Power},
  \citenamefont {Barnes}, \citenamefont {Snyder}, \citenamefont {Schlaggar},\
  and\ \citenamefont {Petersen}}]{power2012spurious}%
  \BibitemOpen
  \bibfield  {author} {\bibinfo {author} {\bibfnamefont {J.~D.}\ \bibnamefont
  {Power}}, \bibinfo {author} {\bibfnamefont {K.~A.}\ \bibnamefont {Barnes}},
  \bibinfo {author} {\bibfnamefont {A.~Z.}\ \bibnamefont {Snyder}}, \bibinfo
  {author} {\bibfnamefont {B.~L.}\ \bibnamefont {Schlaggar}}, \ and\ \bibinfo
  {author} {\bibfnamefont {S.~E.}\ \bibnamefont {Petersen}},\ }\href@noop {}
  {\bibfield  {journal} {\bibinfo  {journal} {Neuroimage}\ }\textbf {\bibinfo
  {volume} {59}},\ \bibinfo {pages} {2142} (\bibinfo {year}
  {2012})}\BibitemShut {NoStop}%
\end{thebibliography}%

\end{document}